\documentclass[11pt]{article}
\usepackage{amsmath,amssymb}
\usepackage[dvipdfmx]{graphicx}
\usepackage{comment}

\begin{document}
\begin{titlepage}

\begin{flushright}
\end{flushright}
\begin{center}
  \vspace{3cm}
  {\bf \Large Cosmological Perturbations via Quantum Corrections \\[0.5cm] in M-theory}
  \\  \vspace{2cm}
  Kazuho Hiraga$^a$ and Yoshifumi Hyakutake$^b$
  \\ \vspace{1cm}
  {${}^a$\it Meijo University Senior High School,  \\
 Shintomi-cho 1-3-16, Nakamura-ku, 
Nagoya, Aichi 453-0031, Japan}\\
   \vspace{0.5cm}
  {${}^b$\it College of Science, Ibaraki University, \\
   Bunkyo 2-1-1, Mito, Ibaraki 310-8512, Japan}
\end{center}

\vspace{2cm}
\begin{abstract}
In the early universe it is important to take into account quantum effect of the gravity to explain the feature of the inflation.
In this paper, we consider the M-theory effective action which consists of 11 dimensional supergravity and (Weyl)$^4$ terms.
The equations of motion are solved perturbatively, and the solution describes the inflation-like expansion in 4 dimensional spacetime.
Scalar and tensor perturbations around this background are evaluated analytically and their behaviors are investigated numerically.
If we assume that these perturbations are constant at the beginning of the inflation, spectral indices for scalar and tensor perturbations
become almost scale invariant.
\end{abstract}

\end{titlepage}


\section{Introduction}\label{sec:intro}


Inflationary expansion of the early universe resolves problems of the hot Big Bang scenario,
such as horizon problem or flatness problem\cite{Starobinsky:1980te}-\cite{Linde:1983gd}. 
It also generates perturbations to the homogeneous background, which become initial conditions 
for the structure formation of the present universe.
Recent observations of cosmological parameters restrict the quantities of these perturbations,
and hence models of the inflation\cite{Ade:2015tva}-\cite{Akrami:2018odb}. 
In order to realize the inflation via 4 dimensional effective theory of the gravity, it is usual to introduce
a inflaton field or some higher curvature terms\footnote{There are a lot of works and reviews in this area.
See refs. \cite{Kolb:1990vq}-\cite{Nojiri:2017ncd}, for examples.}.
The inflaton field causes the inflationary expansion while it slowly rolls down the potential\cite{Linde:1983gd,Freese:1990rb,Linde:1993cn,Boubekeur:2005zm,Bezrukov:2007ep}.
The inflation via higher curvature term was first studied by Starobinsky\cite{Starobinsky:1980te},
which is surprisingly consistent with the current observations.

Since the energy scale of the inflationary era will be near the Planck scale,
it is important to analyze this scenario by using  quantum theory of gravity. 
Actually there have been many attempts to explain the inflation from supergravity theory, 
which is a low energy effective action of superstring theory.
It is difficult, however, to obtain de Sitter (dS) vacua in the effective theory obtained 
by compactification of the supergravity theory, 
which is stated as a no-go theorem\cite{Gibbons:1984kp,Maldacena:2000mw,Gibbons:2003gb}
or dS swampland conjecture\cite{Obied:2018sgi,Garg:2018reu,Ooguri:2018wrx}.
Especially the dS swampland conjecture says that low energy effective theories which realize the inflation by the inflaton field 
cannot be  consistent with quantum theory of gravity at Planck scale.

To evade the no-go theorem, we need to take into account corrections to the supergravity theory.
One way is to introduce sources of branes in the superstring theory, and
it was proposed in ref.~\cite{Kachru:2003aw} that the de Sitter vacua can be constructed 
in the superstring theory with orientifold planes and flux compactification.
Some recent consistency checks with the swampland conjecture are reported in ref.~\cite{Blumenhagen:2020dea}.
Another way is to introduce higher curvature corrections in the superstring theory,
and some earlier works on this direction can be found in refs.~\cite{Ishihara:1986if}-\cite{Akune:2006dg}.

In this paper, we pursue the possibility of the inflationary scenario in M-theory.
The M-theory is defined as a strong coupling limit, or uplift to 11 dimensions, of type IIA superstring theory, 
and the low energy limit is approximated by 11 dimensional supergravity.
Since the type IIA superstring theory contains one-loop corrections, the uplift of these terms
also give corrections to the 11 dimensional supergravity.
As for the metric it gives products of four Weyl tensors, abbreviated as $W^4$, so it is important to understand 
the effect of these terms to the inflationary scenario in the M-theory.
Actually it is shown in ref.~\cite{Hiraga:2018kpb,Hiraga:2019bmp} that if the space-time is divided into 4 dimensions
and 7 internal spatial directions, we obtain a perturbative solution
where 3 spatial directions are expanding and internal ones are shrinking.
Scalar perturbations around the inflationary background are discussed in ref.~\cite{Hiraga:2019syd},
and the scalar spectral index is found to be close to one.
The purpose of this paper is to analyze tensor perturbations around the inflation like background.
We derive an equation of motion for the tensor perturbations, and solve it perturbatively.
We see that both the power spectra of scalar and tensor perturbations are almost scale invariant
if we assume that these are constants at the beginning of the inflation.

Organization of this paper is as follows.
In section 2, we briefly review the background geometry of ref.~\cite{Hiraga:2018kpb}.
In section 3, we consider the tensor perturbations around the background and obtain an analytic expression
for tensor power spectrum perturbatively. The scalar perturbations and the power spectrum of the curvature perturbation
are also summarized.
In section 4, numerical analyses of spectral indices $n_s$ and $n_t$ are discussed.
We also give some comments on the tensor-to-scalar ratio.
Conclusion and discussion are given in section 5.
Since there are many technical calculations in this paper, some of which are done by using Mathematica code,
we collect the results in appendix \ref{sec:supp} and \ref{sec:secS}.


\section{Effective Action and Inflationary Solution in M-theory}\label{sec:review}


Low energy effective action of the superstring theory is obtained by analyzing scattering amplitudes\cite{Gross:1986iv,Gross:1986mw}
or conformal invariance of loop corrections on the string world-sheet\cite{Grisaru:1986px,Grisaru:1986vi}.
The low energy effective action of the M-theory is described by 11 dimensional supergravity,
and its leading correction is obtained by uplifting the results of the type IIA superstring theory.
The bosonic part of the effective action of the M-theory is given by\cite{Tseytlin:2000sf,Becker:2001pm}
\begin{alignat}{3}
  S_{11} &= \frac{1}{2 \kappa_{11}^2} \int d^{11}x \; e \big( R + \Gamma Z \big), \label{eq:W4}
  \\[0.2cm]
  Z &\equiv 24 \big( W_{abcd} W^{abcd} W_{efgh} W^{efgh} - 64 W_{abcd} W^{aefg} W^{bcdh} W_{efgh} \notag
  \\[-0.1cm]
  &\qquad
  + 2 W_{abcd} W^{abef} W^{cdgh} W_{efgh} + 16 W_{acbd} W^{aebf} W^{cgdh} W_{egfh} \notag
  \\
  &\qquad
  - 16 W_{abcd} W^{aefg} W^b{}_{ef}{}^h W^{cd}{}_{gh} - 16 W_{abcd} W^{aefg} W^b{}_{fe}{}^h W^{cd}{}_{gh} \big), \notag
\end{alignat}
where $a,b,c, \cdots =0,1,2, \cdots ,10$ are local Lorentz indices and $W_{abcd}$ is a Weyl tensor\footnote{It is possible 
to make Riemann tensor to Weyl tensor by using field redefinition ambiguity. More general case is discussed in ref.~\cite{Hiraga:2019bmp}.}.
11 dimensional gravitational constant $2\kappa_{11}^2$ and a coefficient $\Gamma$ 
are dimensionful parameters and expressed in terms of 11 dimensional Planck length $\ell_p$ as
\begin{alignat}{3}
  2\kappa_{11}^2 = (2\pi)^8 \ell_p^9, \qquad \Gamma = \frac{\pi^2\ell_p^6}{2^{11} 3^2}. \label{eq:gamma} 
\end{alignat}
Note that the above action can be derived directly in 11 dimensions by imposing local supersymmetry\cite{deRoo:1992sm}-\cite{Hyakutake:2007sm}.
The $Z$ terms in the action~(\ref{eq:W4}) are related to a topological term via local supersymmetry and
the numerical values in this action is protected against higher perturbations.
Of course, there exist higher derivative terms of $\mathcal{O}(\Gamma^n)$ for $n\geq 2$, but those contributions are unknown so far.
Some thoughts on the possible terms of $n=2$ can be found in ref.~\cite{Hiraga:2019syd}.

In this paper, we solve equations of motion perturbatively up to the linear order of $\Gamma$.
By varying the effective action (\ref{eq:W4}), we obtain following equations of motion\cite{Hyakutake:2013vwa}.
\begin{alignat}{3}
  E_{ab} &\equiv R_{ab} - \frac{1}{2} \eta_{ab} R 
  + \Gamma \Big\{ - \frac{1}{2} \eta_{ab} Z + R_{cdea} Y^{cde}{}_b - 2 D_{(c} D_{d)} Y^c{}_{ab}{}^d \Big\} = 0. \label{eq:MEOM}
\end{alignat}
Here $D_a$ is a covariant derivative for local Lorentz index,
and tensors $X_{abcd}$ and $Y_{abcd}$ are defined as
\begin{alignat}{3}
  X_{abcd} \!&=\! 96 \big( W_{abcd} W_{efgh} W^{efgh} \!-\! 4 W_{abce} W_{dfgh} W^{efgh} \!+\! 4 W_{abde} W_{cfgh} W^{efgh} \notag
  \\
  &\quad
  \!-\! 4 W_{cdae} W_{bfgh} W^{efgh} \!+\! 4 W_{cdbe} W_{afgh} W^{efgh} \!+\! 2 W_{abef} W_{cdgh} W^{efgh} \notag
  \\
  &\quad
  \!+\! 4 W_{ab}{}^{ef} W_{ce}{}^{gh} W_{dfgh} \!+\! 4 W_{cd}{}^{ef} W_{ae}{}^{gh} W_{bfgh} \!+\! 8 W_{aecg} W_{bfdh} W^{efgh} \label{eq:Xdef}
  \\
  &\quad
  \!-\! 8 W_{becg} W_{afdh} W^{efgh} \!-\! 8 W_{abeg} W_{cf}{}^e{}_h W_d{}^{fgh} \!-\! 8 W_{cdeg} W_{af}{}^e{}_h W_b{}^{fgh} \notag
  \\
  &\quad
  \!-\! 4 W^{ef}{}_{ag} W_{efch} W^g{}_b{}^h{}_d \!+\! 4 W^{ef}{}_{ag} W_{efdh} W^g{}_b{}^h{}_c \!+\! 4 W^{ef}{}_{bg} W_{efch} W^g{}_a{}^h{}_d \notag
  \\
  &\quad
  \!-\! 4 W^{ef}{}_{bg} W_{efdh} W^g{}_a{}^h{}_c \big), \notag
  \\
  Y_{abcd} &= X_{abcd} - \frac{1}{9} ( \eta_{ac} X_{bd} \!-\! \eta_{bc} X_{ad} \!-\! \eta_{ad} X_{bc} \!+\! \eta_{bd} X_{ac}) 
  + \frac{1}{90} ( \eta_{ac} \eta_{bd} \!-\! \eta_{ad} \eta_{bc} ) X, \label{eq:Ydef}
\end{alignat}
where  $Y^c{}_{acb} = 0$, $ X_{ab} = X^c{}_{acb}$ and $X = X^a{}_a$.

Let us consider the solution of effective action (\ref{eq:W4}) in the early universe.
We assume that the 10 dimensional space directions are divided into 3 dimensional homogeneous space
and 7 dimensional internal one. Then the ansatz of the metric is expressed as
\begin{alignat}{3}
  ds^2 &= - dt^2 + a(t)^2 dx_i^2 + b(t)^2 dy_m^2, \label{eq:bg}
\end{alignat}
where $i=1,2,3$ and $m=4,\cdots,10$.
$a(t)$ and $b(t)$ are scale factors for 3 dimensional space and 7 dimensional internal one, respectively.
By inserting the above ansatz into the eq. (\ref{eq:MEOM}), we obtain differential equations for 
$H(t)=\frac{\dot{a}(t)}{a(t)}$ and $G(t)=\frac{\dot{b}(t)}{b(t)}$.
Here the dot represents the time derivative.
The solution up to linear order of $\Gamma$ is given by\cite{Hiraga:2018kpb}
\begin{alignat}{3}
  &H(\tau) = \frac{H_\text{I}}{\tau} + \frac{c_h \Gamma H_\text{I}^7}{\tau^7} 
  + \mathcal{O} \Big(\frac{\Gamma^2 H_\text{I}^{13}}{\tau^{13}} \Big), \notag
  \\
  &G(\tau) = \frac{-7 + \sqrt{21}}{14} \frac{H_\text{I}}{\tau} + \frac{c_g \Gamma H_\text{I}^7}{\tau^7}
  + \mathcal{O} \Big(\frac{\Gamma^2 H_\text{I}^{13}}{\tau^{13}} \Big), \label{eq:HGsol}
\end{alignat}
with
\begin{alignat}{3}
  &c_h = \frac{13824 (477087 \!-\! 97732\sqrt{21})}{8575} \sim 47111, \notag
  \\
  &c_g = - \frac{41472 (532196 \!-\! 110451 \sqrt{21})}{60025} \sim -17996. \label{eq:chcg}
\end{alignat}
Here $H_\text{I}$ is an integral constant and $\tau$ is defined by
\begin{alignat}{3}
  \tau &= \frac{(-1 + \sqrt{21}) H_\text{I} t + 2}{2}. \label{eq:tau}
\end{alignat}
Notice that $\tau$ is the dimensionless parameter and takes the range of $1 \leq \tau$.
By integrating the eq.~(\ref{eq:HGsol}), scale factor $a(\tau)$ and $b(\tau)$ are written as
\begin{alignat}{3}
  \log \Big(\frac{a}{a_\text{E}}\Big) &= \frac{1+\sqrt{21}}{10} \log \tau 
  - \frac{1+\sqrt{21}}{60} c_h \Gamma H_\text{I}^6 \frac{1}{\tau^6}
  + \mathcal{O} \Big(\frac{\Gamma^2 H_\text{I}^{12}}{\tau^{12}} \Big), \notag
  \\
  \log \Big(\frac{b}{b_\text{E}}\Big) &= - \frac{3\sqrt{21}-7}{70} \log \tau - \frac{1+\sqrt{21}}{60} c_g \Gamma H_\text{I}^6 
  \frac{1}{\tau^6} + \mathcal{O} \Big(\frac{\Gamma^2 H_\text{I}^{12}}{\tau^{12}} \Big), \label{eq:abQc}
\end{alignat}
respectively.
Here $a_\text{E}$ and $b_\text{E} $ are integrate constants and correspond to scale factors around the end of the inflation or the deflation, respectively.


\section{Scalar and Tensor Perturbations in M-theory} \label{sec:scalar}


In this section, we investigate scalar and tensor perturbations around the background geometry (\ref{eq:bg}) 
up to the linear order of $\Gamma$\cite{Hiraga:2019syd}. 
Since the action (\ref{eq:W4}) contains complicated $W^4$ terms, we employ Mathematica code to obtain the results here,
even though these expressions are analytic.
We deal with equations of motion for perturbations order by order in a spirit of ref.~\cite{Weinberg:2008hq},
and auxiliary fields are removed by consulting the case of modified gravity\cite{Hwang:1996xh,DeFelice:2009ak}.

The perturbations around the background metric are chosen as follows.
\begin{alignat}{3}
  ds^2 &= - (1+2\alpha) dt^2 - 2 a \partial_i \beta dt dx^i 
  + a^2 (\delta_{ij} + 2 \partial_i \partial_j \gamma + 2 \psi \delta_{ij} + h_{ij}) dx^i dx^j + b^2 dy_m^2. \label{eq:pt}
\end{alignat}
Here $\alpha$, $\beta$, $\gamma$ and $\psi$ are scalar perturbations and $h_{ij}$ are tensor perturbations.
These perturbations are expanded like 
\begin{alignat}{3}
  &\Upsilon (t,x,y) = \int d^3k d^7l \, \big\{ \Upsilon (t,k,l) e^{i k_i x^i + i l_m y^m} +
  \Upsilon (t,k,l)^\ast e^{-i k_i x^i - i l_m y^m} \big\}, \label{eq:FmPsi}
\end{alignat}
where $\Upsilon$ represents $\alpha$, $\beta$, $\gamma$, $\psi$ or $h_{ij}$.
$k_i$ is a momentum in 3 spatial directions, and $l_m$ is one in 7 internal directions.
It is also possible to include perturbations related to internal space, 
but these appear in the equations of motion with a factor $\frac{l^2}{b^2}$.
Then the perturbations related to internal space decouples from the equations of motion
since $b$ becomes quite small during the inflation\cite{Hiraga:2019syd}. 
Hence we neglect those contributions in this paper.

Linearized equations of motion for the metric perturbations are obtained
by varying the eq.~(\ref{eq:MEOM}). The result is given as follows.
\begin{alignat}{3}
  \delta E_{ab} 
  &= \delta R_{ab} \!-\! \frac{1}{2} \eta_{ab} \delta R 
  + \Gamma \Big\{ \!\!-\! \frac{1}{2} \eta_{ab} \delta R_{cdef} Y^{cdef}
  \!+\! \delta R_{cdea} Y^{cde}{}_b \!+\! R^{cde}{}_a \delta Y_{cdeb} \notag
  \\
  &\quad\,
  - 2 \delta e^\mu{}_c D_\mu D_{d} Y^c{}_{(ab)}{}^d
  - 2 \delta \omega_{c}{}^c{}_e D_{d} Y^e{}_{(ab)}{}^d 
  + 2 \delta \omega_{ce(a} D^{d} Y^{ce}{}_{b)d} \notag
  \\
  &\quad\,
  + 2 \delta \omega_{ce(a} D^{d} Y^c{}_{b)}{}^e{}_d 
  - 2 D_c D_d \delta Y^c{}_{(ab)}{}^d 
  \!-\! 2 D_c \big( \delta e^\nu{}_d D_\nu Y^c{}_{(ab)}{}^d \label{eq:MEOMvar}
  \\
  &\quad\,
  + \delta \omega_d{}^c{}_e Y^e{}_{(ab)}{}^d
  - \delta \omega_{de(a} Y^{ce}{}_{b)}{}^d 
  - \delta \omega_{de(a} Y^c{}_{b)}{}^{ed} 
  + \delta \omega_d{}^d{}_e Y^c{}_{(ab)}{}^e \big) 
  \Big\} 
  = 0. \notag
\end{alignat}
Below we solve the above equations of motion in the momentum space for the tensor perturbations,
and also summarize the results for the scalar perturbations in ref.~\cite{Hiraga:2019syd}.


\subsection{The tensor perturbations}\label{sec:tensor}


In this subsection, we consider the tensor perturbations in 4 dimensional spacetime.
Since the tensor perturbations and scalar perturbations do not mix at the linearized level, we first set
scalar perturbations to zero. Then the metric is given by
\begin{alignat}{3}
 ds^2 = - dt^2 +a^2( \delta_{ij} + h_{ij} ) dx^i dx^j + b^2 dy_m^2,
\end{alignat}
where tensor perturbations $h_{ij}$ can be divided into two polarization modes, $h_+$ and $h_\times$. 
In the vielbein formalism, it is expressed as
\begin{alignat}{3}
  e^a{}_\mu &= \begin{pmatrix} 
    1 & 0 & 0 & 0 & 0 & \cdots & 0 \\
    0 & a(1 \!+\! \frac{1}{2} h_+) & 0 & 0 & 0 & \cdots & 0 \\
    0 & a h_\times & a(1 \!-\! \frac{1}{2} h_+) & 0 & 0 & \cdots & 0 \\
    0 & 0 & 0 & a & 0 & \cdots & 0 \\
    0 & 0 & 0 & 0 & b & \cdots & 0 \\
    \vdots & \vdots & \vdots & \vdots & \vdots & \ddots & 0 \\
    0 & 0 & 0 & 0 & 0 & \cdots & b 
  \end{pmatrix},  \label{eq:tpe}
\end{alignat}
up to the linear order of the perturbation.
By inserting the expression (\ref{eq:tpe}) into the equations of motion (\ref{eq:MEOMvar}), 
we obtain following differential equation for each $h_\alpha (\alpha=+,\times)$.
\begin{alignat}{3}
  0 = \ddot{h}_\alpha + (3H + 7G) \dot{h}_\alpha + \frac{k^2}{a^2} h_\alpha 
  + \Gamma \big( A_0 h_\alpha + A_1 \dot{h}_\alpha  + A_2 \ddot{h}_\alpha + A_3 \dddot{h}_\alpha + A_4 \ddddot{h}_\alpha\big),
  \label{eq:TensorP}
\end{alignat}
where $A_i\, (i=0,1,2,3,4)$ are functions of $H$ and $G$.
Since the explicit forms of $A_i$ are lengthy, we put them in appendix \ref{sec:supp}.
It is also possible to obtain the eq.~(\ref{eq:TensorP}) from an effective action of $\mathcal{O}(\Gamma^2)$, 
and the explicit form is shown in appendix \ref{sec:secS}.
Mathematica codes are located in ref.~\cite{Mathematicacodes2}.

Let us solve the above equation up to the linear order of $\Gamma$. 
First, we expand $H$, $G$ and $h_\alpha$ as 
\begin{alignat}{3}
  H = H_0 + \Gamma H_1, \qquad G = G_0 + \Gamma G_1, \qquad h_\alpha = h_{0} + \Gamma h_{1},
\end{alignat}
and substitute these into the eq.~(\ref{eq:TensorP}) and expand it up to the linear order of $\Gamma$.
Notice that $H_0$ and $G_0$ are leading parts of the background (\ref{eq:HGsol}),
and satisfy $G_0 =\tfrac{-7 +\sqrt{21} }{14}H_0 $ and $\dot{H}_0= \tfrac{1- \sqrt{21} }{2}H_0^2$.
Then the equation of motion for $h_0$ at $\mathcal{O}(\Gamma^0)$ is written as
\begin{alignat}{3}
  0 &= \ddot{h}_{0} + (7 G_0 +3 H_0 ) \dot{h}_{0} + \frac{k^2}{a^2} h_{0} 
  = \ddot{h}_{0} + \frac{1}{2} \big(\sqrt{21}-1\big) H_0 \dot{h}_{0} + \frac{k^2}{a^2}h_{0}. \label{eq:hGamma0th}
\end{alignat}
In order to solve the above equation, we introduce a new time coordinate $\eta$ instead of $t$, 
which is defined by $dt = \frac{1+\sqrt{21}}{10 H_\text{I}} d\tau = a_0 d\eta$.
Note that $a_0 = a_\text{E} \tau^{\frac{1+\sqrt{21}}{10}}$ is the leading part of the scale factor $a$. 
This means that $\eta$ behaves like a conformal time after the inflationary expansion.
Now $\eta$ is expressed in terms of $\tau$ as
\begin{alignat}{3}
  \eta &= \frac{1+\sqrt{21}}{10 H_\text{I}} \int \frac{d\tau}{a_0} 
  = \frac{3+\sqrt{21}}{6 a_\text{E} H_\text{I}} \tau^\frac{9-\sqrt{21}}{10}, \label{eq:conftime}
\end{alignat}
and $a_0$ and a Hubble parameter $\mathcal{H}_0$ with respect to $\eta$ are given by
\begin{alignat}{3}
  a_0 = a_\text{E} \Big( \frac{\sqrt{21}-3}{2} a_\text{E} H_\text{I} \eta \Big)^\frac{3+\sqrt{21}}{6}, \qquad
  \mathcal{H}_0 = \frac{a_0'}{a_0} = \frac{3+\sqrt{21}}{6} \frac{1}{\eta}. \label{eq:a0H0}
\end{alignat}
Here the prime $'$ represent $\frac{d}{d\eta}$.
Then by defining $h_0 = a_0^{\frac{3 - \sqrt{21} }{4}} u_0$ and multiplying $a_0^2$ to the eq.~(\ref{eq:hGamma0th}), we obtain
\begin{alignat}{3}
  0 &= h_0'' +  \frac{\sqrt{21}-3}{2} \mathcal{H}_0 h_0' + k^2 h_0 
  = a_0^{\frac{3-\sqrt{21}}{4}} \bigg[ u_0'' + \Big(k^2 + \frac{1}{4 \eta^2} \Big) u_0 \bigg]. \label{eq:u0}
\end{alignat}
And the above differential equation for $u_0$ can be solved as
\begin{alignat}{3}
  u_0 = c_1 \sqrt{k \eta} J_0(k \eta) + c_2 \sqrt{k \eta} Y_0(k \eta). \label{eq:solu0}
\end{alignat}
Here $J_0$ and $Y_0$ are Bessel functions of the first and second kind, respectively.
$c_1$ and $c_2$ are integral constants which have mass dimension and depend on $k$.
In order to fix the ratio of $\frac{c_2}{c_1}$, we demand that $u_0$ behaves like $e^{-ik\eta}$ as $\eta$ goes to the infinity.
This means that the tensor perturbation is  approximated by the free field as $\eta$ goes to the infinity.
Since $\sqrt{x} J_0(x) \sim \sqrt{\frac{2}{\pi}} \cos(x-\frac{\pi}{4})$ and $\sqrt{x} Y_0(x) \sim \sqrt{\frac{2}{\pi}} \sin(x-\frac{\pi}{4})$
as $x \to \infty$, we choose $\frac{c_2}{c_1} = -i$ and $u_0$ is given by
\begin{alignat}{3}
  u_0 = c_1 \sqrt{k \eta} H^{(2)}_0(k \eta). \label{eq:solu02}
\end{alignat}
$H^{(2)}_0(x)$ is the Hankel function of the second kind. From this, $h_0$ is expressed as
\begin{alignat}{3}
  h_0 &= c_1 a_0^{\frac{3-\sqrt{21} }{4}}  \sqrt{k \eta} H^{(2)}_0(k \eta)
  = \bar{c}_1 H^{(2)}_0(k \eta), \notag
  \\
  \bar{c}_1 &\equiv c_1 a_\text{E}^{\frac{3- \sqrt{21}}{4}} 
  \Big( \frac{\sqrt{21}+3}{6} \frac{k}{a_\text{E} H_\text{I}} \Big)^\frac{1}{2}. \label{eq:solh0}
\end{alignat}
If we take $k\eta \to \infty$, $h_0$ approaches to $\bar{c}_1 \sqrt{\frac{2}{\pi k\eta}} e^{i\frac{\pi}{4}} e^{-ik\eta}$.

Next, we investigate a part of the equation of motion (\ref{eq:TensorP}) which linearly depends on $\Gamma$.
With the aid of Mathematica code\cite{Mathematicacodes2}, we obtain the following equation.
\begin{alignat}{3}
  0 &= \ddot{h}_1 + \frac{1}{2} \big(\sqrt{21}-1\big) \dot{h}_1 H_0 + \frac{ k^2}{a_0^2}h_1 \notag
  \\
  &\quad\,
  + \tfrac{512  (16940714 \sqrt{21}-85692179 )}{8575}  H_0^7   \dot{h}_0
  + \tfrac{256 (613929 \sqrt{21}-2861099 ) }{245} H_0^6 \ddot{h}_0 \notag
  \\
  &\quad\,
  + \tfrac{6144 (121 \sqrt{21}-521)}{7}  H_0^5 \dddot{h}_0 
  +\tfrac{2048 (20 \sqrt{21}-101 )}{7} H_0^4 \ddddot{h}_0 \notag
  \\
  &\quad\,
  + \frac{k^2}{a_0^2} \Big\{ \Big(-2 \bar{a}_1-\tfrac{768 (64897 \sqrt{21}-270367 )}{245}  H_0^6 \Big) h_0
  + \tfrac{1024 (26 \sqrt{21}-383 )}{7} \dot{h}_0 H_0^5 \notag
  \\
  &\quad\,
  + \tfrac{512 (25 \sqrt{21}-43 )}{7}  H_0^4 \ddot{h}_0 \Big\} 
  - \frac{ k^4}{ a_0^4} \tfrac{2048 (7 \sqrt{21}-16 )}{7 } H_0^4 h_0. \label{eq:h1}
\end{alignat}
Here $\bar{a}_1 = - \frac{1+\sqrt{21}}{60} c_h H_0^6$, which comes from $\frac{k^2}{a^2} h_0$.
By multiplying $a_0^2$ to the eq.~(\ref{eq:h1}) and using $\eta$, we obtain
\begin{alignat}{3}
  0 &= h_1'' +\frac{1}{2} \big(\sqrt{21}-3\big) \mathcal{H}_0 h_1' + k^2  h_1 \notag
  \\
  &\quad\,
  +\frac{1}{a_0^6} \Big [ 
  \tfrac{ 256  ( 4950813 \sqrt{21}-33216993 )  }{8575} \mathcal{H}_0^7 h_0' 
  - \tfrac{ 768 (484793-93483 \sqrt{21} ) }{245}   \mathcal{H}_0^6 h_0'' \notag
  \\
  &\quad\,
  + \tfrac{6144 (81 \sqrt{21}-319 ) }{7}   \mathcal{H}_0^5  h_0'''
  - \tfrac{512 (404-80 \sqrt{21} )  }{7}   \mathcal{H}_0^4   h_0'''' \notag
  \\
  &\quad\, 
  + k^2 \Big\{ \big(-\tfrac{  768  (64897 \sqrt{21}-270367\ ) }{245}  \mathcal{H}_0^6-2 a_0^6 \bar{ a }_1 \big)h_0 
  + \tfrac{1536 (9 \sqrt{21}-241) }{7}  \mathcal{H}_0^5 h_0' \notag
  \\
  &\quad\, 
  -\tfrac{512  (43-25 \sqrt{21} ) }{7}  \mathcal{H}_0^4 h_0'' \Big\}
  -k^4  \tfrac{2048  (7 \sqrt{21}-16 )}{7}   \mathcal{H}_0^4 h_0 \Big] . \label{eq:h1eq}
\end{alignat}
In order to solve the above equation, we redefine $h_1$ as $h_1 = a_0^{\frac{3-\sqrt{21}}{4}} u_1$.
Then a differential equation for $u_1$ is expressed as
\begin{alignat}{3}
  0 &= u_1'' +  \Big(k^2-\frac{3(\sqrt{21}-5 ) }{8}  \mathcal{H}_0^2\Big) u_1 \notag
  \\
  &\quad\,
  + \frac{1}{a_0^6} \Big [   \tfrac{  576 (279929593 \sqrt{21}-1282921273 ) }{8575 }  \mathcal{H}_0^8 u_0
  + \tfrac{   4608  (11177551 \sqrt{21}-51407611 )  }{8575}  \mathcal{H}_0^7 u_0' \notag
  \\
  &\quad\,
  - \tfrac{ 768  (1625723-338133 \sqrt{21} ) }{245}  \mathcal{H}_0^6 u_0''
  + \tfrac{8192 (101 \sqrt{21}-420 )  }{7} \mathcal{H}_0^5  u_0'''
  - \tfrac{512(404-80 \sqrt{21} ) }{7}    \mathcal{H}_0^4 u_0'''' \notag
  \\
  &\quad\,
  + k^2\Big\{   \big( -\tfrac{768(55797 \sqrt{21}-234982 ) }{245}   \mathcal{H}_0^6 -2 a_0^6 \bar{a}_1 \big) u_0 
  + \tfrac{1024 (43 \sqrt{21}-525 ) }{7}  \mathcal{H}_0^5 u_0' \notag
  \\
  &\quad\,
  - \tfrac{512(43-25 \sqrt{21} )}{7}   \mathcal{H}_0^4 u_0''      \Big\}
  - k^4 \tfrac{2048  (7 \sqrt{21}-16 ) }{7} \mathcal{H}_0^4  u_0 \Big] \notag
  \\[0.1cm]
  &= u_1'' + \Big( k^2 + \frac{1}{4 \eta ^2} \Big) u_1
  + \frac{ 2^8}{3 \cdot 5^2 \, 7^3} \Big( \frac{\sqrt{21}+3}{6} \Big)^{3+\sqrt{21}}
  \frac{ \sqrt{k \eta } }{a_\text{E}^6 \eta^8 (a_\text{E} H_\text{I} \eta)^{3+\sqrt{21}} } \notag
  \\
  &\quad\,
  \Big[ \big( 73500  (47 \sqrt{21}+217 )  k^3 \eta ^3 
  +6  (1032913 \sqrt{21}+4661457 )  k \eta \big) \big( c_1 J_1(k \eta) + c_2 Y_1(k \eta) \big) \label{eq:u1}
  \\
  &\quad\,
  + \big(-44100 (8 \sqrt{21}+37 )  k^4 \eta^4
  +( 6318421 \sqrt{21}+29265657 ) k^2  \eta^2\big) \big( c_1 J_0(k \eta) + c_2 Y_0(k \eta) \big) \Big]. \notag
\end{alignat}
In the second line, we substituted eqs.~(\ref{eq:a0H0}) and (\ref{eq:solu0}). 
A particular solution for the above equation is given by
\begin{alignat}{3}
  u_1 = - \frac{576 (999- 218 \sqrt{21})}{60025 } 
  \frac{H_{\text{I} }^6 \sqrt{k\eta} }{ \big( \frac{\sqrt{21} - 3}{2}  a_{\text{E} } H_{\text{I} } \eta \big)^{9+\sqrt{21} } } 
  \big( c_1 u_{11} +  \sqrt{\pi}  c_2 u_{12} \big), \label{eq:u1f}
\end{alignat}
and the explicit forms of $u_{11}$ and $u_{12}$ are shown in the appendix \ref{sec:supp}.
The ratio of $\frac{c_2}{c_1}$ should be fixed by $\frac{c_2}{c_1} = -i$ as explained before.
Thus we have solved the eq.~(\ref{eq:h1eq}) as
\begin{alignat}{3}
  h_1 &= - \frac{576 (999- 218 \sqrt{21})}{60025 } c_1 a_0^{\frac{3-\sqrt{21} }{4} } 
  \frac{ H_{\text{I} }^6 \sqrt{k\eta} }{ \big( \frac{\sqrt{21} - 3}{2}  a_{\text{E} } H_{\text{I} } \eta \big)^{9+\sqrt{21} } } 
  \big( u_{11} - i   \sqrt{\pi}  u_{12} \big) \notag
  \\
  &= - \frac{576 (999- 218 \sqrt{21})}{60025 } \bar{c}_1 \frac{  H_{\text{I} }^6  }{\tau^6 } 
  \big( u_{11} - i   \sqrt{\pi}  u_{12} \big). \label{eq:solh1}
\end{alignat}
Here we used $\tau^6 = ( \frac{\sqrt{21}-3}{2} a_\text{E} H_\text{I} \eta )^{9+\sqrt{21}}$.


\subsection{The scalar perturbations}


In this subsection, we briefly summarize the results of scalar perturbations obtained in ref.~\cite{Hiraga:2019syd}.
The metric for the scalar perturbations are written as
\begin{alignat}{3}
  ds^2 &= - (1+2\alpha) dt^2 - 2 a \partial_i \beta dt dx^i 
  + a^2 (\delta_{ij} + 2 \partial_i \partial_j \gamma + 2 \psi \delta_{ij}) dx^i dx^j + b^2 dy_m^2. \label{eq:pts}
\end{alignat}
By inserting the above into the eq.~(\ref{eq:MEOMvar}), we obtain four independent equations with respect to
$\alpha$, $\chi \equiv a(\beta + a \dot{\gamma})$ and $P \equiv H^{-1}\psi$.
By using two equations out of four, it is possible to solve $\alpha$ and $\chi$ up to the linear order of $\Gamma$,
and one equation becomes redundant.
Then we obtain a differential equation for $P = P_0 + \Gamma P_1$ up to the linear order of $\Gamma$. 
\begin{alignat}{3}
  0 &= \ddot{P}_0 - \frac{\sqrt{21}-1}{2} H_0 \dot{P}_0 
  + \Big( \frac{k^2}{a_0^2} - \frac{\sqrt{21}-11}{2} H_0^2 \Big) P_0 \notag
  \\
  &\quad\,
  +\Gamma
   \Big[ \ddot{P}_1 - \frac{\sqrt{21}-1}{2} H_0 \dot{P}_1 
  + \Big( \frac{k^2}{a_0^2} - \frac{\sqrt{21}-11}{2} H_0^2 \Big) P_1 \notag
  \\
  &\quad\,
  - \tfrac{1536 (49692383 \sqrt{21}-70593438) }{8575} H_0^8 P_0
  + \tfrac{768 (36412229 \sqrt{21}-124991079) }{1715} H_0^7 \dot{P}_0 \notag
  \\
  &\quad\,
  + \tfrac{768 (5604373\sqrt{21}-36068337) }{1715} H_0^6 \ddot{P}_0
  + \tfrac{12288 (6383 \sqrt{21}-17688) }{245} H_0^5 \dddot{P}_0 \label{eq:P}
  \\
  &\quad\,
  + \tfrac{3072 (2261\sqrt{21}-23271) }{1225} H_0^4 \ddddot{P}_0
  + \frac{k^2}{a_0^2} \Big\{ \Big( \tfrac{768 (3567079\sqrt{21}-29260239) }{8575} H_0^6 
  - 2 \bar{a}_1 \Big) P_0 \notag
  \\
  &\quad\,
  + \tfrac{3072 (85331 \sqrt{21}-265416) }{1225} H_0^5 \dot{P}_0
  + \tfrac{1536 (9479 \sqrt{21}-66369) }{1225} H_0^4 \ddot{P}_0 \Big\} 
  + \frac{k^4}{a_0^4} \tfrac{6144 (1633 \sqrt{21}-9288) }{1225} H_0^4 P_0 \Big]. \notag
\end{alignat}
Here $\bar{a}_1 = - \frac{1+\sqrt{21}}{60} c_h H_0^6$, which comes from $\frac{k^2}{a^2} P_0$.
As in the case of the tensor perturbations, in order to solve the eq.~(\ref{eq:P}), 
we use the variable $\eta$ which is defined in the eq.~(\ref{eq:conftime}). 
By multiplying $a_0^2$ to the eq.~(\ref{eq:P}), we obtain
\begin{alignat}{3}
  0 &= P_0'' - \frac{\sqrt{21}+1}{2} \mathcal{H}_0 P_0' + \Big( k^2 - \frac{\sqrt{21}-11}{2} \mathcal{H}_0^2 \Big) P_0 \notag
  \\
  &\quad\, 
  + \Gamma \bigg[ P_1'' - \frac{\sqrt{21}+1}{2} \mathcal{H}_0 P_1' 
  + \Big( k^2 - \frac{\sqrt{21}-11}{2} \mathcal{H}_0^2 \Big) P_1 \notag
  \\
  &\quad\, 
  + \frac{1}{a_0^6} \Big\{ -\tfrac{1536 (49692383 \sqrt{21}-70593438) }{8575} \mathcal{H}_0^8 P_0 
  + \tfrac{1536 (15053494 \sqrt{21}-40585737) }{1715} \mathcal{H}_0^7 P_0' \notag
  \\
  &\quad\, 
  + \tfrac{6912 (1812421 \sqrt{21}-16802761) }{8575} \mathcal{H}_0^6 P_0''
  + \tfrac{6144 (57047 \sqrt{21}-107067) }{1225} \mathcal{H}_0^5 P'''_0 \label{eq:P1-2}
  \\
  &\quad\, 
  + \tfrac{3072 (2261 \sqrt{21}-23271) }{1225} \mathcal{H}_0^4 P''''_0
  + k^2 \big\{ \big( \tfrac{768 (3567079 \sqrt{21}-29260239) }{8575} \mathcal{H}_0^6 - 2 a_0^6 \bar{a}_1 \big) P_0 \notag
  \\
  &\quad\, 
  + \tfrac{1536 (161183 \sqrt{21}-464463) }{1225} \mathcal{H}_0^5 P_0'
  + \tfrac{1536 (9479 \sqrt{21}-66369) }{1225} \mathcal{H}_0^4 P_0'' \big\} \notag
  \\
  &\quad\,
  + k^4 \tfrac{6144 (1633 \sqrt{21}-9288) }{1225} \mathcal{H}_0^4 P_0 \Big\} \bigg]. \notag 
\end{alignat}

Let us solve the above equation perturbatively.
First, by setting $P_0 = a_0^{\frac{\sqrt{21}+1}{4}} U_0$, a part of $\mathcal{O}(\Gamma^0)$ becomes
\begin{alignat}{3}
  0 &= P_0'' - \frac{\sqrt{21}+1}{2} \mathcal{H}_0 P_0' + \Big( k^2 - \frac{\sqrt{21}-11}{2} \mathcal{H}_0^2 \Big) P_0 \notag
  \\
  &= a_0^{\frac{\sqrt{21}+1}{4}} \bigg[ U_0'' + \Big(k^2 + \frac{1}{4 \eta^2} \Big) U_0 \bigg]. \label{eq:U0}
\end{alignat}
This is the same as the eq.~(\ref{eq:u0}). 
Then $U_0$, $P_0$ and $\psi_0 = H_0 P_0$ are solved as follows.
\begin{alignat}{3}
  U_0 &= c_1 \sqrt{k \eta} H^{(2)}_0(k \eta), \notag
  \\
  P_0 &= c_1 a_0^{\frac{\sqrt{21}+1}{4}} \sqrt{k \eta} H^{(2)}_0(k \eta), \label{eq:solU0etc}
  \\
  \psi_0 &= \tilde{c}_1 H^{(2)}_0(k \eta), \qquad
  \tilde{c}_1 = c_1 a_\text{E}^{\frac{\sqrt{21}+1}{4}} H_\text{I} 
  \Big( \frac{\sqrt{21}+3}{6} \frac{k}{a_\text{E} H_\text{I}} \Big)^{\frac{1}{2} }. \notag
\end{alignat}
In the above, we set $\frac{c_2}{c_1}=-i$ as in the tensor perturbations.
In the limit of $k\eta \to \infty$, $\psi_0(\eta,k)$ approaches to 
$\tilde{c}_1 \sqrt{\frac{2}{\pi k\eta}} e^{i\frac{\pi}{4}} e^{-ik\eta}$.

Next let us solve a part of the differential equation (\ref{eq:P1-2}) which linearly depends on $\Gamma$.
By setting $P_1 = a_0^{\frac{\sqrt{21}+1}{4}} U_1$ and using the solution (\ref{eq:solU0etc}), 
the equation which linearly depends on $\Gamma$ becomes
\begin{alignat}{3}
  0 &= U_1'' + \Big( k^2 + \frac{1}{4 \eta ^2} \Big) U_1
  + \frac{ 2^8}{3 \cdot 5^2 \, 7^3} \Big( \frac{\sqrt{21}+3}{6} \Big)^{3+\sqrt{21}}
  \frac{\sqrt{k \eta}}{a_\text{E}^6 \eta^8 (a_\text{E} H_\text{I} \eta)^{3+\sqrt{21}} } \notag
  \\
  &\quad\,
  \Big[ \big( 420 (127267+27753 \sqrt{21}) k^3 \eta^3
  - 78 (27149229+5923661 \sqrt{21}) k \eta \big) \big( c_1 J_1(k \eta) + c_2 Y_1(k \eta) \big) \notag
  \\
  &\quad\,
  + \big( -44100 (37+8\sqrt{21}) k^4 \eta^4 + (602616417+131422261 \sqrt{21}) k^2 \eta^2, \label{eq:U1}
  \\
  &\quad\,
  + 36 (49428849+10780291 \sqrt{21}) \big) \big( c_1 J_0(k \eta) + c_2 Y_0(k \eta) \big) \Big]. \notag
\end{alignat}
A particular solution of the above is obtained by using Mathematica code\cite{Hiraga:2019syd}.
The solutions of $U_1$, $P_1$ and $\psi_1 = H_1 P_0 + H_0 P_1 = H_0 (c_h H_0^6 P_0 + P_1)$ are expressed as follows.
\begin{alignat}{3}
  U_1 &= - \tfrac{288 (20727 - 4523 \sqrt{21})}{300125} 
  \frac{H_\text{I}^6 \sqrt{k \eta}}{\big( \frac{\sqrt{21}-3}{2} a_\text{E} H_\text{I} \eta \big)^{9+\sqrt{21}}} 
  \big( c_1 U_{11} - c_2 \tfrac{(- 41 + 9 \sqrt{21}) \sqrt{\pi}}{10} U_{12} \big), \notag
  \\
  P_1 &= - \tfrac{288 (20727 - 4523 \sqrt{21})}{300125} c_1 a_0^{\frac{\sqrt{21}+1}{4}}
  \frac{H_\text{I}^6 \sqrt{k \eta}}{\big( \frac{\sqrt{21}-3}{2} a_\text{E} H_\text{I} \eta \big)^{9+\sqrt{21}}} 
  \big( U_{11} + i \tfrac{(- 41 + 9 \sqrt{21}) \sqrt{\pi}}{10} U_{12} \big), \label{eq:solU1etc}
  \\
  \psi_1 &= \frac{ \tilde{c}_1 H_\text{I}^6 }{ \tau^6 } \Big\{ c_h H^{(2)}_0(k \eta) 
  - \tfrac{288 (20727 - 4523 \sqrt{21})}{300125} 
  \big( U_{11} + i \tfrac{(- 41 + 9 \sqrt{21}) \sqrt{\pi}}{10} U_{12} \big) \Big\}. \notag
\end{alignat}
The explicit expressions of $U_{11}$ and $U_{12}$ can be found in ref.~\cite{Hiraga:2019syd}.
Here, the ratio of integration constants is fixed as $\frac{c_2}{c_1} = -i$.
We also used $\tau^6 = ( \frac{\sqrt{21}-3}{2} a_\text{E} H_\text{I} \eta )^{9+\sqrt{21}}$ 
and $H_1= c_h H_0 ^6$ from the eq.~(\ref{eq:HGsol}).
Note that as $\tau$ approaches to the infinity, $\psi_1$ decreases faster than $\psi_0$.


\section{Numerical Analyses for Scalar and Tensor Perturbations}

\begin{figure}[htb]
 \centering
 \begin{picture}(270,180)
 \put(272,5){$\bar{\eta}$}
 \put(0,168){$\frac{1}{aH}$}
 \includegraphics[keepaspectratio]{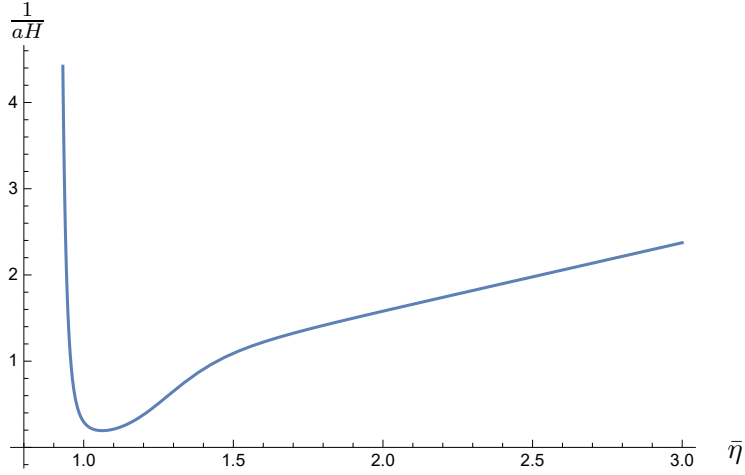}
 \end{picture}
 \caption{Plot of the comoving Hubble radius $\frac{1}{aH}$ as a function of $\bar{\eta}$.}
 \label{fig:comovingHR}
\end{figure}

In this section, we examine spectral indices of the scalar and tensor perturbations.
First of all, we change the definition of $\tau$ by rescaling.
Without loss of generality, it is possible to rescale $\tau$ like
\begin{alignat}{3}
  \tau \;\to\; (c_h \Gamma H_\text{I}^6)^\frac{1}{6} \tau.
\end{alignat}
And after this prescription, we shift the integral constants as
\begin{alignat}{3}
  H_\text{I} \;\to\; (c_h \Gamma H_\text{I}^6)^\frac{1}{6} H_\text{I}, \quad
  a_\text{E} \;\to\; (c_h \Gamma H_\text{I}^6)^{-\frac{1+\sqrt{21}}{60}} a_\text{E}, \quad
  b_\text{E} \;\to\; (c_h \Gamma H_\text{I}^6)^{\frac{3\sqrt{21}-7}{420}} b_\text{E}.
\end{alignat}
Then $\eta$, $a_0$, $b_0$, $H_0$ and $G_0$ are invariant, and $H(\tau)$, $G(\tau)$, $a(\tau)$ and $b(\tau)$ behave as
\begin{alignat}{3}
  \frac{H}{H_\text{I}} &= \frac{1}{\tau} + \frac{1}{\tau^7} + \mathcal{O}\Big(\frac{1}{\tau^{13}}\Big), \notag
  \\
  \frac{G}{H_\text{I}} &= \frac{-7 + \sqrt{21}}{14} \frac{1}{\tau}
  + \frac{c_g}{c_h} \frac{1}{\tau^7} + \mathcal{O}\Big(\frac{1}{\tau^{13}}\Big), \label{eq:HGab}
  \\
  \log \Big(\frac{a}{a_\text{E}}\Big) &= \frac{1+\sqrt{21}}{10} \log \tau 
  - \frac{1+\sqrt{21}}{60} \frac{1}{\tau^6}
  + \mathcal{O} \Big(\frac{1}{\tau^{12}} \Big), \notag
  \\
  \log \Big(\frac{b}{b_\text{E}}\Big) &= - \frac{3\sqrt{21}-7}{70} \log \tau 
  - \frac{1+\sqrt{21}}{60} \frac{c_g}{c_h} \frac{1}{\tau^6} 
  + \mathcal{O} \Big(\frac{1}{\tau^{12}} \Big). \notag
\end{alignat}
Note that the range of $\tau$ becomes $(c_h \Gamma H_\text{I}^6)^{-\frac{1}{6}} \leq \tau$. 
Thus the parameter $\Gamma H_\text{I}^6$ disappears from the background and absorbed into the lower bound of $\tau$,
which is determined by requiring that the e-folding number $N_\text{e}$ is within the range of $60 < N_\text{e} < 70$.
Below we use the eq.~(\ref{eq:HGab}) as the background and the range of $\tau$ is given by $\tau_\text{I} < \tau$.
The explicit value of $\tau_\text{I}$ is irrelevant in this paper but should be determined by the e-folding 
number\footnote{For example, $N_\text{e} = 69$ corresponds to $\tau_\text{I} \sim 0.34$ ($\bar{\eta} \sim 0.78$)\cite{Hiraga:2019syd}.
Higher derivative terms which are not considered in this paper will also affect the explicit value of $\tau_\text{I}$.}.
From the eq.~(\ref{eq:conftime}), $\eta$ is also bounded as
\begin{alignat}{3}
  \frac{3+\sqrt{21}}{6} \tau_\text{I}^\frac{9-\sqrt{21}}{10} \leq a_\text{E} H_\text{I} \eta \equiv \bar{\eta}.
\end{alignat}
Here we introduced dimensionless parameter $\bar{\eta}$, and
$\tau=1.0$ corresponds to $\bar{\eta} = \frac{3+\sqrt{21}}{6} \sim 1.3$.
Below we assume that the expansions in the eq.~(\ref{eq:HGab}) are convergent around 
$\tau\sim 0 \, (\bar{\eta} \sim 0)$ and coefficients of higher order terms in the eq.(\ref{eq:HGab}) are small and 
negligible around $\tau \sim 0.50\, (\bar{\eta} \sim 0.93)$\footnote{
Some thoughts on this point is given in the appendix of ref.~\cite{Hiraga:2019syd}.}.
A plot of the comoving Hubble radius $\frac{1}{aH}$ as a function of $\bar{\eta}$ 
is shown in fig.~\ref{fig:comovingHR}.
From this figure, we see that the inflation ends around $\tau \sim 0.7\, (\bar{\eta} \sim 1.1)$.

\begin{figure}[htb]
 \centering
 \begin{picture}(280,185)
 \put(275,155){$\bar{\eta}$}
 \put(-55,5){$\log \frac{\mathcal{P}_t(\bar{\eta},\bar{k})}{\mathcal{P}_t(0.9,\bar{k})}$}
 \includegraphics[keepaspectratio, scale=1]{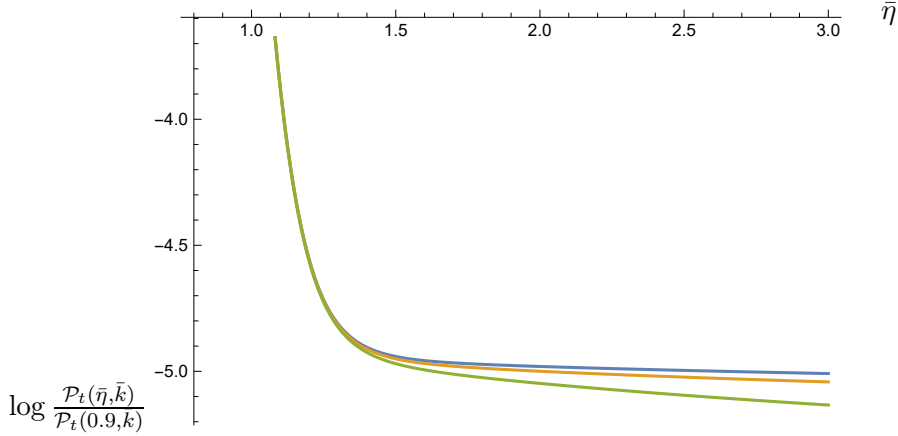}
 \end{picture}
 \caption{Plots of $\log \frac{\mathcal{P}_t(\bar{\eta},\bar{k})}{\mathcal{P}_t(0.9,\bar{k})}$ with $\bar{k}=e^{-30}, e^{-20}$ and $e^{-10}$
 (from top to bottom).}
 \label{fig:TensorPt1}
\end{figure}

\begin{figure}[htb]
 \centering
 \begin{picture}(280,185)
 \put(-30,158){$\log \bar{k}$}
 \put(270,5){$\log \frac{\mathcal{P}_t(3.0,\bar{k})}{\mathcal{P}_t(0.9,\bar{k})}$}
 \includegraphics[keepaspectratio, scale=1]{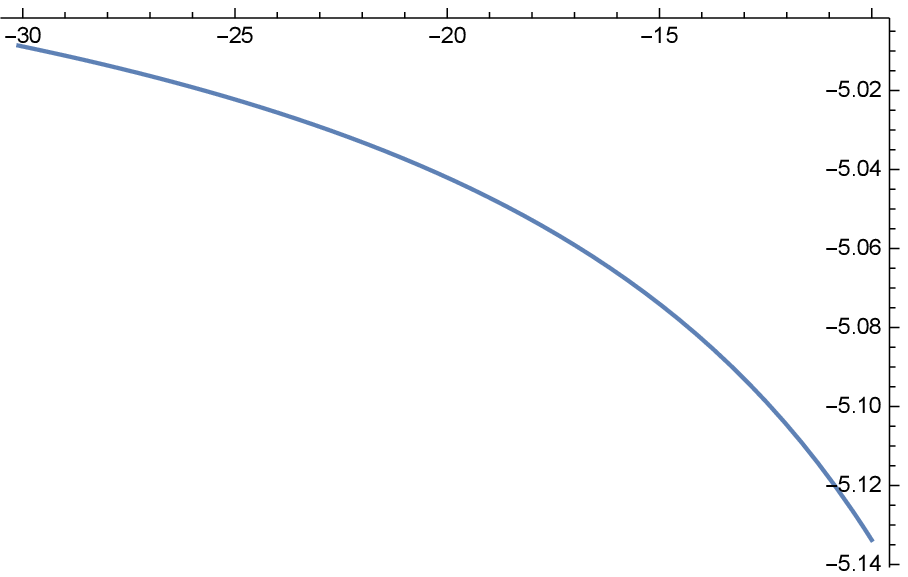}
 \end{picture}
 \caption{Plots of $\log \frac{\mathcal{P}_t(3.0,\bar{k})}{\mathcal{P}_t(0.9,\bar{k})}$
 as a function of $\bar{k}$.}
 \label{fig:TensorPt2}
\end{figure}

Let us evaluate the power spectra of scalar and tensor perturbations.
First, from the eqs.~(\ref{eq:solh0}) and (\ref{eq:solh1}), the power spectrum of the tensor perturbations is expressed as
\begin{alignat}{3}
  \mathcal{P}_t(\bar{\eta},\bar{k}) &= \bar{k}^3 | h |^2 \notag
  \\
  &= |\bar{c}_1|^2 \bar{k}^3 \Big| H_0^{(2)}(\bar{k} \bar{\eta}) 
  - \frac{576 (999- 218 \sqrt{21})}{60025 \, c_h \tau^6} 
  \big( u_{11} - i \sqrt{\pi} u_{12} \big) \Big|^2, \label{eq:PST}
\end{alignat}
where $\bar{k} \equiv \frac{k}{a_\text{E} H_\text{I}}$ is a dimensionless momentum\footnote{A sum of physical degrees of freedom $\alpha=+,\times$
are included in $\bar{c}_1$}.
Since $\bar{c}_1$ cannot be determined, we normalize the power spectrum as 
$\frac{\mathcal{P}_t(\bar{\eta},\bar{k})}{\mathcal{P}_t(0.9,\bar{k})}$.
Plots of these functions with $\bar{k} = e^{-30}, e^{-20}$ and $e^{-10}$ 
are shown in fig.~\ref{fig:TensorPt1}.
Naively the horizon exit occurs at $k=aH$, but the corrections should modify this relation.
Thus we define the spectral index $n_t$ by evaluating the power spectrum at $\bar{\eta}=3.0$,
where the scale factor behaves like a radiation dominated era.
A function of $\log \frac{\mathcal{P}_t(3.0,\bar{k})}{\mathcal{P}_t(0.9,\bar{k})}$ is plotted in fig.~\ref{fig:TensorPt2}.
If we fit the curve, we obtain
\begin{alignat}{3}
  \log \frac{\mathcal{P}_t(3.0,\bar{k})}{\mathcal{P}_t(0.9,\bar{k})}
  &= - 5.4 - 0.033 \log \bar{k} - 0.0011 (\log \bar{k})^2 - 1.4 \times 10^{-5} (\log \bar{k})^3. \label{eq:Tensorfit}
\end{alignat}
From this, we see that the tensor spectral index becomes $n_t = - 0.033$ and its runnings are quite small.
Thus if the power spectrum is independent on $k$ at the beginning of the inflation,
it is almost scale independent after the inflation.
Note that if we evaluate $n_t$ at $\bar{\eta}=2.0$, it becomes closer to zero.

\begin{figure}[htb]
 \centering
 \begin{picture}(280,185)
 \put(270,125){$\bar{\eta}$}
 \put(-55,5){$\log \frac{\mathcal{P}_s(\bar{\eta},\bar{k})}{\mathcal{P}_s(0.9,\bar{k})}$}
 \includegraphics[keepaspectratio, scale=1]{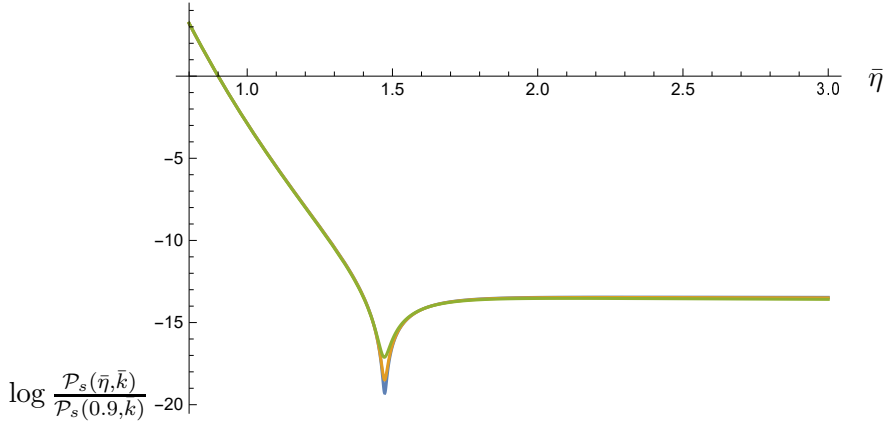}
 \end{picture}
 \caption{Plots of $\log \frac{\mathcal{P}_s(\bar{\eta},\bar{k})}{\mathcal{P}_s(0.9,\bar{k})}$ with 
 $\bar{k} = e^{-30}, e^{-20}$ and $e^{-10}$ (from bottom to top at $\bar{\eta}=1.5$).}
 \label{fig:ScalarPt1}
\end{figure}

\begin{figure}[htb]
 \centering
 \begin{picture}(280,185)
 \put(-30,155){$\log \bar{k}$}
 \put(270,5){$\log \frac{\mathcal{P}_s(3.0,\bar{k})}{\mathcal{P}_s(0.9,\bar{k})}$}
 \includegraphics[keepaspectratio, scale=1]{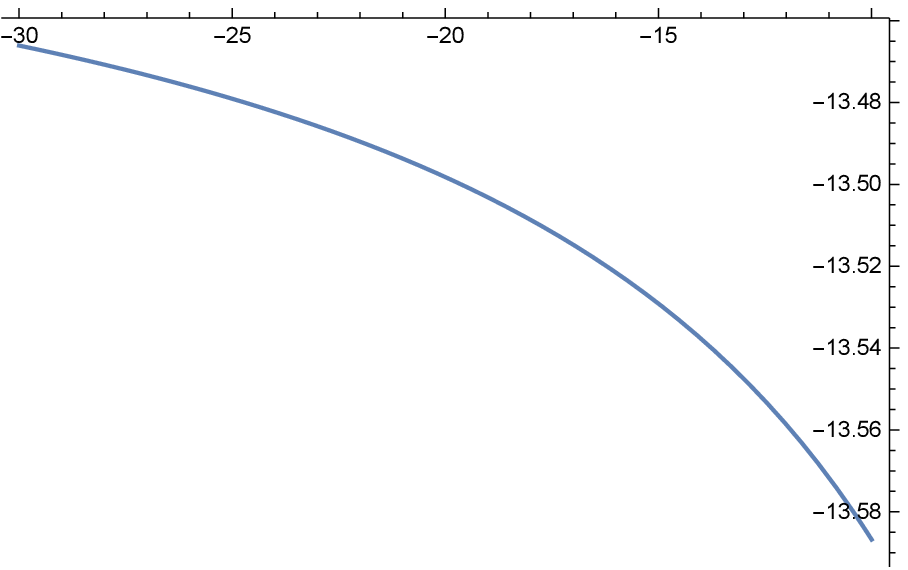}
 \end{picture}
 \caption{Plots of $\log \frac{\mathcal{P}_s(3.0,\bar{k})}{\mathcal{P}_s(0.9,\bar{k})}$
 as a function of $\bar{k}$.}
 \label{fig:ScalarPt2}
\end{figure}

Next, from the eqs.~(\ref{eq:solU0etc}) and (\ref{eq:solU1etc}), 
the power spectrum of the scalar perturbations is expressed as
\begin{alignat}{3}
  \mathcal{P}_s (\bar{\eta},\bar{k}) &= \bar{k}^3 | \psi |^2 \label{eq:PSS}
  \\
  &= |\tilde{c}_1|^2 \bar{k}^3 \Big| 
  \Big(1 \!\!+\!\! \frac{1}{\tau^6}\Big) H_0^{(2)}(k\eta) 
  \!-\! \frac{288 (20727 \!-\! 4523 \sqrt{21})}{300125 \, c_h \tau^6} 
  \Big( U_{11} \!-\! i \frac{(41 \!-\! 9 \sqrt{21}) \sqrt{\pi}}{10} U_{12} \Big) \Big|^2. \notag
\end{alignat}
Since the dimensionful parameter $\tilde{c}_1$ cannot be determined, we normalize the power spectrum as 
$\frac{\mathcal{P}_s(\bar{\eta},\bar{k})}{\mathcal{P}_s(0.9,\bar{k})}$.
Plots of these functions with $\bar{k} = e^{-30}, e^{-20}$ 
and $e^{-10}$ are shown in fig.~\ref{fig:ScalarPt1}.
And a function of $\log \frac{\mathcal{P}_s(3.0,\bar{k})}{\mathcal{P}_s(0.9,\bar{k})}$ is plotted in fig.~\ref{fig:ScalarPt2}.
If we fit the data, we obtain
\begin{alignat}{3}
  \log \frac{\mathcal{P}_s(3.0,\bar{k})}{\mathcal{P}_s(0.9,\bar{k})}
  &= - 14 - 0.032 \log \bar{k} - 0.0011 (\log \bar{k})^2 - 1.4 \times 10^{-5} (\log \bar{k})^3. \label{eq:Scalarfit}
\end{alignat}
From this, we see that the spectral index becomes $n_s = 0.97$ and its runnings are quite small.
Again if the power spectrum is independent on $k$ at the beginning of the inflation,
it is almost scale independent after the inflation.
Note that if we evaluate $n_s$ at $\bar{\eta}=2.0$, it becomes closer to one.

The tensor-to-scalar ratio $r$ is obtained by the power spectra of scalar and tensor perturbations.
If we define $r$ at $\bar{k}=e^{-10}$, we obtain
\begin{alignat}{3}
  r &=\frac{\mathcal{P}_t(3.0,e^{-10})}{\mathcal{P}_s(3.0,e^{-10})} 
  = \frac{\mathcal{P}_t(3.0,e^{-10})}{\mathcal{P}_t(0.9,e^{-10})} 
  \frac{\mathcal{P}_s(0.9,e^{-10})}{\mathcal{P}_s(3.0,e^{-10})} 
  \frac{\mathcal{P}_t(0.9,e^{-10})}{\mathcal{P}_s(0.9,e^{-10})}
  = e^{8.5} \frac{\mathcal{P}_t(0.9,e^{-10})}{\mathcal{P}_s(0.9,e^{-10})}.
  \label{eq:TtoS}
\end{alignat}
Here we used $\log \frac{\mathcal{P}_t(3.0,e^{-10})}{\mathcal{P}_t(0.9,e^{-10})}=-5.1$
and $\log \frac{\mathcal{P}_s(3.0,e^{-10})}{\mathcal{P}_s(0.9,e^{-10})}=-13.6$.
Since $r < 0.06$, we need that the power spectrum of the tensor perturbation is much smaller
than that of the scalar perturbation at the beginning of the inflation\cite{Akrami:2018odb}.
The mechanism of this process is not clear so far, and the knowledge of the behaviors of perturbations 
around $\bar{\eta} \sim 0$ should be quite important.


\section{Conclusion and Discussion}


In this paper, we examined the inflationary solution via higher derivative corrections in the M-theory 
and examined scalar and tensor perturbations around such background.
As a result, we have obtained power spectra of these perturbations analytically up to the linear order 
of $\Gamma$, and found that these are almost scale independent after the inflation.
Since we do not know explicit forms of higher order terms, we simply neglect them throughout the paper.

We analyzed the inflationary solution, and scalar and tensor perturbations numerically in section 4.
The background geometry is given by the eq.~(\ref{eq:HGab}) up to the linear order of $\Gamma$,
and the behavior of the comoving Hubble radius is shown in the fig.~\ref{fig:comovingHR}. 
From this we see that the inflation is realized during $\bar{\eta}<1.1$, and
the lower bound for $\bar{\eta}$ should be determined by the e-folding number.
In the numerical analyses, we assumed that the expression (\ref{eq:HGab}) is valid for $0.9 < \bar{\eta}$.

The tensor perturbations $h_\alpha = h_0 + \Gamma h_1$ are evaluated in the background of (\ref{eq:HGab}).
The equations of motion for $h_\alpha$ are solved perturbatively. The solution of $h_0$ is given by the 
eq.~(\ref{eq:solh0}), and that of $h_1$ is analytically obtained by the eq.~(\ref{eq:solh1}).
The power spectrum of the tensor perturbations is evaluated as in the eq.~(\ref{eq:PST}),
and the numerical plot is shown in the fig.~\ref{fig:TensorPt1}.
If we fit the plot up to $(\log \bar{k})^3$ at $\bar{\eta}=3.0$, the function is approximately given by the eq.~(\ref{eq:Tensorfit}).
This implies that the spectral index becomes $n_t = -0.033$, and its runnings are quite small.
Thus if the power spectrum is independent on $k$ at the beginning of the inflation,
it is almost scale independent after the inflation.

In a similar way, curvature perturbation $\psi = \psi_0 + \Gamma \psi_1$ is evaluated 
in the background of (\ref{eq:HGab}).
The equation of motion for $\psi$ is perturbatively solved in the ref.~\cite{Hiraga:2019syd}. 
The solution for $\psi_0$ is given by the eq.~(\ref{eq:solU0etc}), and that of $\psi_1$ 
is done by the eq.~(\ref{eq:solU1etc}).
The power spectrum of the scalar perturbation is expressed as the eq.~(\ref{eq:PSS}),
and the numerical plots are shown in the fig.~\ref{fig:ScalarPt1}.
The plots take minimum values around $\bar{\eta}=1.5$, but become almost constants after $\bar{\eta}=2.0$.
If we fit the plot up to $(\log \bar{k})^3$ at $\bar{\eta}=3.0$, the function is approximately obtained 
as the eq.~(\ref{eq:Scalarfit}).
This implies that the spectral index becomes $n_s = 0.97$, and its runnings are quite small.
Again if the power spectrum is independent on $k$ at the beginning of the inflation,
it is almost scale independent after the inflation.

Thus we conclude that the scalar and tensor spectral indices are almost scale invariant in the M-theory inflation,
if the power spectra are independent on $k$ at the beginning of the inflation.
On the other hand, the tensor-to-scalar ratio is estimated as the eq.~(\ref{eq:TtoS}).
Since $r < 0.06$, the power spectrum of the tensor perturbation should be much smaller
than that of the scalar perturbation at the beginning of the inflation\cite{Akrami:2018odb}.
In order to explain this problem at least we need to know the behaviors of perturbations around $\bar{\eta} \sim 0$.

As a future work, it is interesting to apply the method developed here 
to more complicated internal geometry, such as $G_2$ manifold \cite{Brandhuber:2001yi}. 
It is also interesting to apply the analyses of this paper 
to the heterotic superstring theory with nontrivial internal space, which contains $R^2$ corrections\cite{Brandle:2000qp},
and reveal several problems in string cosmology\cite{Antoniadis:2016avv}. 
Unification of the inflationary expansion and late time acceleration in modified gravity, such as $f(R)$ gravity or mimetic gravity, is an interesting direction 
to be explored\cite{Nojiri:2003rz}-\cite{Sebastiani:2016ras}.


\section*{Acknowledgement}


The authors would like to thank Takanori Fujiwara and Makoto Sakaguchi.
This work was partially supported by Japan Society for the Promotion of Science, 
Grant-in-Aid for Scientific Research (C) Grant Number JP17K05405.

\appendix

\section{Supplementary Notes} \label{sec:supp}


In this appendix, we show the explicit forms of $A_i\, (i=0,1,2,3,4)$ in the eq.~(\ref{eq:TensorP}),
and $u_{11}$ and $u_{12}$ in the eq.~(\ref{eq:u1f}).
The results are obtained by using the Mathematica codes, which are located in \cite{Mathematicacodes2}.

First, the forms of $A_i$ in the eq.~(\ref{eq:TensorP}) are given by 
\begin{alignat}{3}
  A_0 &= \tfrac{3584\, k^2}{10125\, a^2} \big( 164 G^6-156051 G^5 H+552087 G^4 H^2+126177 G^4 \dot{H}-169557 \dot{G} G^4 \notag
  \\
  &\quad\,
  - 634247 G^3 H^3 +42420 G^3 \ddot{H}+383772 \dot{G} G^3 H-227652 G^3 H \dot{H}-45450 G^3 \ddot{G} \notag
  \\
  &\quad\,
  + 237309 G^2 H^4 - 3501 \dot{G} G^2 H^2 -145749 G^2 H^2 \dot{H}+131070 G^2 H \ddot{G}-118800 G^2 H \ddot{H} \notag
  \\
  &\quad\,
  + 3030 G^2 \dddot{H} -21708 G^2 \dot{H}^2 +39906 \dot{G} G^2 \dot{H}-3030 G^2 \dddot{G}-15018 \dot{G}^2 G^2-954 G H^5 \notag
  \\
  &\quad\,
  + 29592 \dot{G} H^4 -10710 H^3 \ddot{G} -240306 \dot{G} G H^3+243966 G H^3 \dot{H}-74910 G H^2 \ddot{G} \notag
  \\
  &\quad\,
  + 59460 G H^2 \ddot{H} -6210 H^2 \dddot{G} -30111 \dot{G}^2 H^2-12588 \dot{G} H^2 \dot{H}-24720 G \dot{H} \ddot{G} \notag
  \\
  &\quad\,
  - 27900 \dot{G} G \ddot{H} +21540 G \dot{H} \ddot{H}  +26160 \dot{G} H \ddot{G}-32520 H \dot{H} \ddot{G}-29340 \dot{G} H \ddot{H}
  \\
  &\quad\,
  - 6360 \ddot{G} \ddot{H} +9240 G H \dddot{G} -9240 G H \dddot{H}-3180 \dot{H} \dddot{G}-3180 \dot{G} \dddot{H}+185049 \dot{G}^2 G H \notag
  \\
  &\quad\,
  + 115749 G H \dot{H}^2  -307158 \dot{G} G H \dot{H}-10737 \dot{G} \dot{H}^2-6753 \dot{G}^2 \dot{H}+31080 \dot{G} G \ddot{G} \notag
  \\
  &\quad\,
  + 3180 \ddot{G}^2 +3180 \dot{G} \dddot{G}+8081 \dot{G}^3+1692 H^6+3258 H^4 \dot{H}+16920 H^3 \ddot{H}+6210 H^2 \dddot{H} \notag
  \\
  &\quad\,
  + 45879 H^2 \dot{H}^2+3180 \ddot{H}^2+35700 H \dot{H} \ddot{H}+3180 \dot{H} \dddot{H}+9409 \dot{H}^3 \big) \notag
  \\
  &\quad\,
  - \tfrac{14336 \,k^4 }{675 \,a^4} \big( 92 G^4-322 G^3 H+401 G^2 H^2-46 G^2 \dot{H}+46 \dot{G} G^2-204 G H^3+72 \dot{G} H^2\notag
  \\
  &\quad\,
  - 118 \dot{G} G H  +118 G H \dot{H} +26 \dot{G} \dot{H}-13 \dot{G}^2+33 H^4-72 H^2 \dot{H}-13 \dot{H}^2 \big). \notag
\end{alignat}
\vspace{-0.8cm}
\begin{alignat}{3}
  A_1 &= \tfrac{3584}{3375} \big( 56 G^7 + 82155 H G^6 - 286878 H^2 G^5 + 88465 \dot{G} G^5 - 63454 \dot{H} G^5 + 243124 H^3 G^4 \notag
  \\
  &\quad\,
  - 83314 H \dot{G} G^4 + 26141 \ddot{G} G^4 + 3967 H \dot{H} G^4-24251 \ddot{H} G^4 + 33174 H^4 G^3 + 119992 \dot{G}^2 G^3 \notag
  \\
  &\quad\,
  + 80084 \dot{H}^2 G^3 - 369471 H^2 \dot{G} G^3 - 52736 H \ddot{G} G^3 + 369900 H^2 \dot{H} G^3 - 199896 \dot{G} \dot{H} G^3 \notag
  \\
  &\quad\,
  + 45356 H \ddot{H} G^3 + 1930 \dddot{G} G^3 - 1930 \dddot{H} G^3 - 61587 H^5 G^2 - 213335 H \dot{G}^2 G^2 \notag
  \\
  &\quad\,
  - 37713 H \dot{H}^2 G^2 + 303948 H^3 \dot{G} G^2 - 25977 H^2 \ddot{G} G^2 + 39778 \dot{G} \ddot{G} G^2 - 180903 H^3 \dot{H} G^2 \notag
  \\
  &\quad\,
  + 241268 H \dot{G} \dot{H} G^2 - 37578 \ddot{G} \dot{H} G^2 + 32157 H^2 \ddot{H} G^2 - 38678 \dot{G} \ddot{H} G^2 + 36478 \dot{H} \ddot{H} G^2 \notag
  \\
  &\quad\,
  - 4690 H \dddot{G} G^2 + 4690 H \dddot{H} G^2 - 7776 H^6 G + 19057 \dot{G}^3 G - 25346 \dot{H}^3 G - 124134 H^2 \dot{G}^2 G \notag
  \\
  &\quad\,
  + 2020 \ddot{G}^2 G - 192414 H^2 \dot{H}^2 G + 75909 \dot{G} \dot{H}^2 G + 2020 \ddot{H}^2 G + 55728 H^4 \dot{G} G \notag
  \\
  &\quad\,
  + 43998 H^3 \ddot{G} G - 86454 H \dot{G} \ddot{G} G - 110772 H^4 \dot{H} G - 69620 \dot{G}^2 \dot{H} G + 335568 H^2 \dot{G} \dot{H} G \notag
  \\
  &\quad\,
  + 69734 H \ddot{G} \dot{H} G - 41778 H^3 \ddot{H} G + 78094 H \dot{G} \ddot{H} G - 4040 \ddot{G} \ddot{H} G - 61374 H \dot{H} \ddot{H} G \notag
  \\
  &\quad\,
  + 510 H^2 \dddot{G} G + 2020 \dot{G} \dddot{G} G - 2020 \dot{H} \dddot{G} G - 510 H^2 \dddot{H} G - 2020 \dot{G} \dddot{H} G + 2020 \dot{H} \dddot{H} G \notag
  \\
  &\quad\,
  - 2268 H^7 - 24056 H \dot{G}^3 - 25095 H \dot{H}^3 + 17277 H^3 \dot{G}^2 - 5100 H \ddot{G}^2 - 50157 H^3 \dot{H}^2 
  \\
  &\quad\,
  + 19974 H \dot{G} \dot{H}^2 + 7841 \ddot{G} \dot{H}^2 - 5100 H \ddot{H}^2 + 4644 H^5 \dot{G} + 8574 H^4 \ddot{G} + 1681 \dot{G}^2 \ddot{G} \notag
  \\
  &\quad\,
  - 21084 H^2 \dot{G} \ddot{G} - 18738 H^5 \dot{H} + 29177 H \dot{G}^2 \dot{H} + 23460 H^3 \dot{G} \dot{H} + 35604 H^2 \ddot{G} \dot{H} \notag
  \\
  &\quad\,
  - 9522 \dot{G} \ddot{G} \dot{H} - 11484 H^4 \ddot{H} - 3221 \dot{G}^2 \ddot{H} - 9381 \dot{H}^2 \ddot{H} + 28344 H^2 \dot{G} \ddot{H} + 10200 H \ddot{G} \ddot{H} \notag
  \\
  &\quad\,
  - 42864 H^2 \dot{H} \ddot{H} + 12602 \dot{G} \dot{H} \ddot{H} + 2250 H^3 \dddot{G} - 5100 H \dot{G} \dddot{G} + 5100 H \dot{H} \dddot{G} - 2250 H^3 \dddot{H} \notag
  \\
  &\quad\,
  + 5100 H \dot{G} \dddot{H} - 5100 H \dot{H} \dddot{H} \big) + \tfrac{7168\, k^2}{675\,a^2} \big( 259 G^5 + 863 G^4 H - 3235 G^3 H^2 - 1744 G^3 \dot{H} \notag
  \\
  &\quad\,
  + 2010 \dot{G} G^3 + 2713 G^2 H^3 - 266 G^2 \ddot{H} - 1746 \dot{G} G^2 H + 1142 G^2 H \dot{H} + 266 G^2 \ddot{G} - 468 G H^4 \notag
  \\
  &\quad\,
  + 528 \dot{G} H^3 + 72 H^2 \ddot{G} - 792 \dot{G} G H^2 + 1202 G H^2 \dot{H} - 338 G H \ddot{G} + 338 G H \ddot{H} - 194 \dot{H} \ddot{G} \notag
  \\
  &\quad\,
  - 194 \dot{G} \ddot{H} + 1017 G \dot{H}^2 - 2228 \dot{G} G \dot{H} - 241 \dot{G}^2 H + 288 \dot{G} H \dot{H} + 194 \dot{G} \ddot{G} + 1211 \dot{G}^2 G \notag
  \\
  &\quad\,
  - 132 H^5 - 600 H^3 \dot{H} - 72 H^2 \ddot{H} + 194 \dot{H} \ddot{H} - 47 H \dot{H}^2 \big). \notag
\end{alignat}
\vspace{-0.8cm}
\begin{alignat}{3}
  A_2 &= - \tfrac{3584}{3375} \big( 1952 G^6 + 27187 G^5 H + 30271 G^4 H^2 - 23579 G^4 \dot{H} + 34969 \dot{G} G^4 - 112291 G^3 H^3 \notag
  \\
  &\quad\,
  - 10190 G^3 \ddot{H} + 163296 \dot{G} G^3 H - 138316 G^3 H \dot{H} + 11110 G^3 \ddot{G} + 21957 G^2 H^4 \notag
  \\
  &\quad\,
  - 50623 \dot{G} G^2 H^2 - 11037 G^2 H^2 \dot{H} + 29850 G^2 H \ddot{G} - 29530 G^2 H \ddot{H} - 920 G^2 \dddot{H} \notag
  \\
  &\quad\,
  + 44556 G^2 \dot{H}^2 - 139692 \dot{G} G^2 \dot{H} + 920 G^2 \dddot{G} + 98216 \dot{G}^2 G^2 + 24228 G H^5 - 19524 \dot{G} H^4 \notag
  \\
  &\quad\,
  - 15690 H^3 \ddot{G} - 128118 \dot{G} G H^3 + 130938 G H^3 \dot{H} - 25270 G H^2 \ddot{G} + 21870 G H^2 \ddot{H} 
  \\
  &\quad\,
  - 2160 H^2 \dddot{G} + 7367 \dot{G}^2 H^2 - 68874 \dot{G} H^2 \dot{H} - 40460 G \dot{H} \ddot{G} - 43540 \dot{G} G \ddot{H} + 37380 G \dot{H} \ddot{H} \notag
  \\
  &\quad\,
  + 27300 \dot{G} H \ddot{G} - 33460 H \dot{H} \ddot{G} - 30380 \dot{G} H \ddot{H} - 6160 \ddot{G} \ddot{H} + 1240 G H \dddot{G} - 1240 G H \dddot{H} \notag
  \\
  &\quad\,
  - 3080 \dot{H} \dddot{G} - 3080 \dot{G} \dddot{H} + 130037 \dot{G}^2 G H + 126477 G H \dot{H}^2 - 262674 \dot{G} G H \dot{H} - 8001 \dot{G} \dot{H}^2 \notag
  \\
  &\quad\,
  - 15099 \dot{G}^2 \dot{H} + 46620 \dot{G} G \ddot{G} + 3080 \ddot{G}^2 + 3080 \dot{G} \dddot{G} + 12733 \dot{G}^3 + 6696 H^6 + 41994 H^4 \dot{H} \notag
  \\
  &\quad\,
  + 17850 H^3 \ddot{H} + 2160 H^2 \dddot{H} + 64587 H^2 \dot{H}^2 + 3080 \ddot{H}^2 + 36540 H \dot{H} \ddot{H} + 3080 \dot{H} \dddot{H} \notag
  \\
  &\quad\,
  + 10367 \dot{H}^3 \big)
  + \tfrac{7168\, k^2}{675\,a^2} \big( 37 G^4 + 118 G^3 H - 479 G^2 H^2 - 266 G^2 \dot{H} + 266 \dot{G} G^2 + 456 G H^3 \notag
  \\
  &\quad\,
  + 72 \dot{G} H^2 - 338 \dot{G} G H + 338 G H \dot{H} - 194 \dot{G} \dot{H} + 97 \dot{G}^2 - 132 H^4 - 72 H^2 \dot{H} + 97 \dot{H}^2 \big). \notag
\end{alignat}
\vspace{-0.8cm}
\begin{alignat}{3}
  A_3 &= -\tfrac{28672}{675} \big( 14 G^5 + 272 G^4 H - 229 G^3 H^2 - 284 G^3 \dot{H} + 330 \dot{G} G^3 - 315 G^2 H^3 \notag
  \\
  &\quad\,
  - 46 G^2 \ddot{H} + 686 \dot{G} G^2 H - 670 G^2 H \dot{H} + 46 G^2 \ddot{G} + 159 G H^4 - 348 \dot{G} H^3 \notag
  \\
  &\quad\,
  - 108 H^2 \ddot{G} - 668 \dot{G} G H^2 + 498 G H^2 \dot{H} + 62 G H \ddot{G} - 62 G H \ddot{H} - 154 \dot{H} \ddot{G} 
  \\
  &\quad\,
  - 154 \dot{G} \ddot{H} + 477 G \dot{H}^2 - 1108 \dot{G} G \dot{H} + 293 \dot{G}^2 H - 740 \dot{G} H \dot{H} + 154 \dot{G} \ddot{G} \notag
  \\
  &\quad\,
  + 631 \dot{G}^2 G + 99 H^5 + 456 H^3 \dot{H} + 108 H^2 \ddot{H} + 154 \dot{H} \ddot{H} + 447 H \dot{H}^2 \big). \notag 
\end{alignat}
\vspace{-0.8cm}
\begin{alignat}{3}
  A_4 &= -\tfrac{14336}{675} \big( 2 G^4 + 38 G^3 H - 49 G^2 H^2 - 46 G^2 \dot{H} + 46 \dot{G} G^2 - 24 G H^3 - 108 \dot{G} H^2 \notag
  \\
  &\quad\,
  + 62 \dot{G} G H - 62 G H \dot{H} - 154 \dot{G} \dot{H} + 77 \dot{G}^2 + 33 H^4 + 108 H^2 \dot{H} + 77 \dot{H}^2 \big).
\end{alignat}

Next, the explicit form of $u_{11}$ in the eq.(\ref{eq:u1f}) is given by  
\begin{alignat}{3}
  u_{11} &= 7 \sqrt{\frac{\pi}{k\eta}} J_0(k \eta)
  \Big\{ - 14700 (146 \sqrt{21} + 669 ) k^3 \eta^3 G_{3,5}^{2,2} \Big( k \eta , \tfrac{1}{2} \Big|
  \begin{array}{c}
    \tfrac{5}{4},\tfrac{2 \sqrt{21}+17 }{4} ,\tfrac{1}{4} \\
    \tfrac{3}{4},\tfrac{3}{4},\frac{1}{4},\tfrac{3}{4},\tfrac{2 \sqrt{21}+13}{4} \\
  \end{array} \Big) \notag
  \\
  &\quad\,
  + 24500 ( 857 \sqrt{21} + 3927 ) k^2 \eta^2 G_{3,5}^{2,2} \Big( k \eta , \tfrac{1}{2} \Big|
  \begin{array}{c}
    \tfrac{5}{4},\tfrac{2 \sqrt{21}+19 }{4}  ,\tfrac{3}{4} \\
    \tfrac{5}{4},\tfrac{5}{4},\frac{1}{4},\tfrac{1}{4},\tfrac{ 2 \sqrt{21}+15}{4} \\
  \end{array} \Big) \notag
  \\
  &\quad\,
  + ( 38465701 \sqrt{21} + 176254865 ) k \eta G_{3,5}^{2,2} \Big( k \eta ,\tfrac{1}{2} \Big|
  \begin{array}{c}
    \tfrac{5}{4},\tfrac{2 \sqrt{21}+21 }{4},\tfrac{1}{4} \\
    \tfrac{3}{4},\tfrac{3}{4},\tfrac{1}{4},\tfrac{3}{4},\tfrac{2 \sqrt{21}+17}{4} \\
  \end{array} \Big) \notag
  \\
  &\quad\,
  + 6 ( 6206377 \sqrt{21} + 28445153 ) G_{3,5}^{2,2} \Big(k \eta  ,\tfrac{1}{2} \Big|
  \begin{array}{c}
    \tfrac{5}{4},\tfrac{2 \sqrt{21}+23 }{4}  ,\tfrac{3}{4} \\
    \tfrac{5}{4},\tfrac{5}{4},\tfrac{1}{4},\tfrac{1}{4},\tfrac{2 \sqrt{21}+19}{4} \\
  \end{array} \Big) \Big\} \label{eq:u11}
  \\
  &\quad\,
  + \pi  ( k \eta)^2 Y_0(k \eta )
  \Big\{ - 51450 (61 \sqrt{21} + 279)  k^2 \eta^2 \, 
  _2 F_3 \big( \tfrac{1}{2},-\tfrac{\sqrt{21}}{2}-\tfrac{5}{2};1,1,-\tfrac{\sqrt{21}}{2}-\tfrac{3}{2};-k^2 \eta^2 \big) \notag
  \\
  &\quad\,
  + 85750 \big(179 \sqrt{21}+819\big) k^2 \eta^2  \,
  _2F_3\big(\tfrac{3}{2},-\tfrac{\sqrt{21}}{2}-\tfrac{5}{2};2,2,-\tfrac{\sqrt{21}}{2}-\tfrac{3}{2};-k^2 \eta^2\big) \notag
  \\
  &\quad\,
  + (46502521 \sqrt{21} + 213002167) \, 
  _2F_3 \big( \tfrac{1}{2},-\tfrac{\sqrt{21}}{2} -\tfrac{7}{2}; 1 , 1,- \tfrac{\sqrt{21}}{2} - \tfrac{5}{2}; -k^2 \eta ^2 \big) \notag
  \\
  &\quad\,
  + 3 (7499743 \sqrt{21} + 34391077) \,
  _2F_3\big(\tfrac{3}{2},-\tfrac{\sqrt{21}}{2}-\tfrac{7}{2};2,2,-\tfrac{\sqrt{21}}{2}-\tfrac{5}{2};-k^2 \eta^2\big) \Big\}. \notag
\end{alignat}
Here the function 
$G_{p,q}^{m,n} \big(z,r|\begin{array}{c} a_1, \cdots, a_n, a_{n+1}, \cdots, a_p \\ b_1, \cdots, b_m, b_{m+1}, \cdots, b_q \\ \end{array} \big)$
is the generalized Meijer G-function, and the function 
$_p F_q \big(a_1,\cdots,a_p; b_1,\cdots,b_q; z \big)$ is the generalized hypergeometric function.
Finally, the explicit form of $u_{12}$ in the eq.(\ref{eq:u1f}) is given by  
\begin{alignat}{3}
  u_{12} &= \sqrt{k\eta} J_0(k \eta )
  \Big\{ 51450 \big(61 \sqrt{21}+279 \big) \sqrt{\pi }  ( k \eta )^{\frac{7}{2}}\, 
  _2F_3 \big(\tfrac{1}{2},-\tfrac{\sqrt{21}}{2}-\tfrac{5}{2};1,1,-\tfrac{\sqrt{21}}{2}-\tfrac{3}{2};-k^2 \eta^2 \big) \notag
  \\
  &\quad\,
  - 85750 \big(179 \sqrt{21} + 819 \big) \sqrt{\pi}  ( k \eta )^{\frac{7}{2}} \, 
  _2F_3 \big( \tfrac{3}{2},-\tfrac{\sqrt{21}}{2}-\tfrac{5}{2};2,2,-\tfrac{\sqrt{21}}{2}-\tfrac{3}{2};-k^2 \eta^2 \big)  \notag
  \\
  &\quad\,
  - \big( 46502521 \sqrt{21} + 213002167 \big) \sqrt{\pi} ( k \eta )^{\frac{3}{2}} \, 
  _2F_3\big(\tfrac{1}{2},-\tfrac{\sqrt{21}}{2}-\tfrac{7}{2};1,1,-\tfrac{\sqrt{21}}{2}-\tfrac{5}{2};-k^2 \eta^2 \big)  \notag
  \\
  &\quad\,
  - 3 \big(7499743 \sqrt{21}+ 34391077\big) \sqrt{\pi } ( k \eta  )^{\frac{3}{2}} \, 
  _2F_3 \big(\tfrac{3}{2},-\tfrac{\sqrt{21}}{2}-\tfrac{7}{2};2,2,-\tfrac{\sqrt{21}}{2}-\tfrac{5}{2};-k^2 \eta^2 \big)  \notag
  \\
  &\quad\,
  - 205800 (146 \sqrt{21} + 669) k^3 \eta^3 G_{3,5}^{3,1} \Big(k \eta ,\tfrac{1}{2} \Big|
  \begin{array}{c}
    \frac{2 \sqrt{21}+15}{4} ,\frac{3}{4},\frac{3}{4} \\
    \frac{1}{4},\frac{1}{4},\frac{1}{4},\frac{3}{4},\frac{2 \sqrt{21}+11}{4}    \\
  \end{array} \Big)  \notag
  \\
  &\quad\,
  + 343000 \big(857 \sqrt{21}+3927\big)k^2 \eta ^2  G_{3,5}^{3,1}\Big(k \eta ,\tfrac{1}{2} \Big|
  \begin{array}{c}
    \frac{2 \sqrt{21}+17}{4}   ,\frac{1}{4},\frac{1}{4} \\
    -\frac{1}{4},-\frac{1}{4},\frac{3}{4},\frac{1}{4},\frac{2 \sqrt{21}+13}{4}  \\
  \end{array} \Big) \notag
  \\
  &\quad\,
  + 14 \big(38465701 \sqrt{21}+176254865 \big) k \eta  G_{3,5}^{3,1} \Big(k \eta ,\tfrac{1}{2} \Big|
  \begin{array}{c}
    \frac{2 \sqrt{21}+19}{4}  ,\frac{3}{4},\frac{3}{4} \\
    \frac{1}{4},\frac{1}{4},\frac{1}{4},\frac{3}{4},\frac{2 \sqrt{21}+15}{4}   \\
  \end{array} \Big) \notag
  \\
  &\quad\,
  + 84 \big(6206377 \sqrt{21}+28445153 \big) G_{3,5}^{3,1} \Big(k \eta ,\tfrac{1}{2} \Big|
  \begin{array}{c}
    \frac{2 \sqrt{21}+21}{4}  ,\frac{1}{4},\frac{1}{4} \\
    -\frac{1}{4},-\frac{1}{4},\frac{3}{4},\frac{1}{4},\frac{2 \sqrt{21}+17}{4}  \\
  \end{array} \Big) \Big\} \label{eq:u12}
  \\
  &\quad\,  
  + \frac{7}{\sqrt{k\eta}}  Y_0(k \eta) 
  \Big\{ 14700 (146 \sqrt{21} + 669 ) k^3 \eta^3  G_{3,5}^{2,2}\Big(k \eta ,\tfrac{1}{2} \Big|
  \begin{array}{c}
    \frac{5}{4},\frac{2 \sqrt{21}+17}{4}  ,\frac{1}{4} \\
    \frac{3}{4},\frac{3}{4},\frac{1}{4},\frac{3}{4},\frac{2 \sqrt{21}+13}{4} \\
  \end{array} \Big) \notag
  \\
  &\quad\,
  - 24500 \big(857 \sqrt{21}+3927\big) k^2 \eta ^2  G_{3,5}^{2,2}\Big(k \eta ,\tfrac{1}{2} \Big|
  \begin{array}{c}
    \frac{3}{4},\frac{2 \sqrt{21}+19}{4}  ,-\frac{1}{4} \\
    \frac{1}{4},\frac{5}{4},-\frac{1}{4},\frac{1}{4},\frac{2 \sqrt{21}+15}{4}  \\
  \end{array} \Big) \notag
  \\
  &\quad\,
  - \big(38465701 \sqrt{21}+176254865\big) k \eta  G_{3,5}^{2,2}\Big(k \eta ,\tfrac{1}{2} \Big|
  \begin{array}{c}
    \frac{5}{4},\frac{2 \sqrt{21}+21}{4}   ,\frac{1}{4} \\
    \frac{3}{4},\frac{3}{4},\frac{1}{4},\frac{3}{4},\frac{2 \sqrt{21}+17}{4}  \\
  \end{array} \Big) \notag
  \\
  &\quad\,
  - 6 \big(6206377 \sqrt{21}+28445153\big) G_{3,5}^{2,2}\Big(k \eta ,\tfrac{1}{2} \Big| 
  \begin{array}{c}
    \frac{3}{4},\frac{2 \sqrt{21}+23}{4}  ,-\frac{1}{4} \\
    \frac{1}{4},\frac{5}{4},-\frac{1}{4},\frac{1}{4},\frac{2 \sqrt{21}+19}{4}  \\
  \end{array} \Big) \Big\}.  \notag
\end{alignat}


\section{Effective Action for the Tensor Perturbation} \label{sec:secS}


In this appendix, we show an effective action of the tensor perturbations up to the second order.
The results are obtained by using the Mathematica code\cite{Mathematicacodes2}.
The effective action for $h_\times$ is obtained as
\begin{alignat}{3}
  S_{\text{pt} }^{(2)} 
  &= \frac{1}{2 \kappa_{11}^2} \int d^{11}x \ a^3 b^7 
  \Big[ \frac{1}{2} \dot{h}_\times^2 - \frac{k^2}{2 a^2}  h_\times^2
  + \Gamma \Big( B_0 h_\times^2 + B_1 \dot{h}_\times^2 + B_2 \ddot{h}_\times^2 \Big) \Big], \label{eq:deltasecond}
\end{alignat}
where explicit forms of $B_i\, (i=0,1,2)$ in the eq.(\ref{eq:deltasecond}) are given by  
\begin{alignat}{3}
  B_0 &= - \tfrac{1792\, k^2 }{10125\, a^2} \big( 
  164 G^6 - 156051 G^5 H + 552087 G^4 H^2 + 126177 G^4 \dot{H} - 169557 \dot{G} G^4 \notag
  \\
  &\quad\,
  - 634247 G^3 H^3 + 42420 G^3 \ddot{H} + 383772 \dot{G} G^3 H - 227652 G^3 H \dot{H} - 45450 G^3 \ddot{G} \notag
  \\
  &\quad\,
  + 237309 G^2 H^4 - 3501 \dot{G} G^2 H^2 - 145749 G^2 H^2 \dot{H} + 131070 G^2 H \ddot{G} - 118800 G^2 H \ddot{H} \notag
  \\
  &\quad\,
  + 3030 G^2 \dddot{H} - 21708 G^2 \dot{H}^2 + 39906 \dot{G} G^2 \dot{H} - 3030 G^2 \dddot{G} - 15018 \dot{G}^2 G^2 - 954 G H^5 \notag
  \\
  &\quad\,
  + 29592 \dot{G} H^4 - 10710 H^3 \ddot{G} - 240306 \dot{G} G H^3 + 243966 G H^3 \dot{H} - 74910 G H^2 \ddot{G} \notag
  \\
  &\quad\,
  + 59460 G H^2 \ddot{H} - 6210 H^2 \dddot{G} - 30111 \dot{G}^2 H^2 - 12588 \dot{G} H^2 \dot{H} - 24720 G \dot{H} \ddot{G} \notag
  \\
  &\quad\,
  - 27900 \dot{G} G \ddot{H} + 21540 G \dot{H} \ddot{H} + 26160 \dot{G} H \ddot{G} - 32520 H \dot{H} \ddot{G} - 29340 \dot{G} H \ddot{H} 
  \\
  &\quad\,
  - 6360 \ddot{G} \ddot{H} + 9240 G H \dddot{G} - 9240 G H \dddot{H} - 3180 \dot{H} \dddot{G} - 3180 \dot{G} \dddot{H} + 185049 \dot{G}^2 G H \notag
  \\
  &\quad\,
  + 115749 G H \dot{H}^2 - 307158 \dot{G} G H \dot{H} - 10737 \dot{G} \dot{H}^2 - 6753 \dot{G}^2 \dot{H} + 31080 \dot{G} G \ddot{G} \notag
  \\
  &\quad\,
  + 3180 \ddot{G}^2 + 3180 \dot{G} \dddot{G} + 8081 \dot{G}^3 + 1692 H^6 + 3258 H^4 \dot{H} + 16920 H^3 \ddot{H} \notag
  \\
  &\quad\,
  + 6210 H^2 \dddot{H} + 45879 H^2 \dot{H}^2 + 3180 \ddot{H}^2 + 35700 H \dot{H} \ddot{H} + 3180 \dot{H} \dddot{H} + 9409 \dot{H}^3 \big) \notag
  \\
  &\quad\,
  + \tfrac{7168\, k^4 }{675\, a^4} \big( 92 G^4 - 322 G^3 H + 401 G^2 H^2 - 46 G^2 \dot{H} + 46 \dot{G} G^2 - 204 G H^3 \notag
  \\
  &\quad\,
  + 72 \dot{G} H^2 - 118 \dot{G} G H + 118 G H \dot{H} + 26 \dot{G} \dot{H} - 13 \dot{G}^2 + 33 H^4 - 72 H^2 \dot{H} - 13 \dot{H}^2 \big). \notag
\end{alignat}
\vspace{-0.8cm}
\begin{alignat}{3}
  B_1 &= \tfrac{1792}{3375} \big( 8 G^6+11733 G^5 H-46011 G^4 H^2-10741 G^4 \dot{H}+12631 \dot{G} G^4+54451 G^3 H^3 \notag
  \\
  &\quad\,
  - 1930 G^3 \ddot{H} - 25696 \dot{G} G^3 H + 18316 G^3 H \dot{H} + 1930 G^3 \ddot{G} - 18597 G^2 H^4 \notag
  \\
  &\quad\, 
  - 15477 \dot{G} G^2 H^2 + 21657 G^2 H^2 \dot{H} - 4690 G^2 H \ddot{G} + 4690 G^2 H \ddot{H} + 8824 G^2 \dot{H}^2 \notag
  \\
  &\quad\,
  - 18748 \dot{G} G^2 \dot{H} + 9924 \dot{G}^2 G^2 - 828 G H^5 + 1824 \dot{G} H^4 + 2250 H^3 \ddot{G} + 26718 \dot{G} G H^3 \notag
  \\
  &\quad\, 
  - 24498 G H^3 \dot{H} + 510 G H^2 \ddot{G} - 510 G H^2 \ddot{H} - 3147 \dot{G}^2 H^2 + 13554 \dot{G} H^2 \dot{H} \notag
  \\
  &\quad\,
  - 2020 G \dot{H} \ddot{G} - 2020 \dot{G} G \ddot{H} + 2020 G \dot{H} \ddot{H} - 5100 \dot{G} H \ddot{G} + 5100 H \dot{H} \ddot{G} \notag
  \\
  &\quad\,
  + 5100 \dot{G} H \ddot{H} - 23717 \dot{G}^2 G H - 15357 G H \dot{H}^2 + 39074 \dot{G} G H \dot{H} + 2741 \dot{G} \dot{H}^2
  \\
  &\quad\,
  - 1201 \dot{G}^2 \dot{H} + 2020 \dot{G} G \ddot{G} - 113 \dot{G}^3 - 756 H^6 - 4734 H^4 \dot{H} - 2250 H^3 \ddot{H} \notag
  \\
  &\quad\,
  - 10407 H^2 \dot{H}^2 - 5100 H \dot{H} \ddot{H} - 1427 \dot{H}^3 \big) \notag
  \\
  &\quad\,
  + \tfrac{3584\, k^2}{675\, a^2}  \big( 37 G^4 + 118 G^3 H - 479 G^2 H^2 - 266 G^2 \dot{H} + 266 \dot{G} G^2 + 456 G H^3 \notag
  \\
  &\quad\,
  +72 \dot{G} H^2-338 \dot{G} G H  +338 G H \dot{H}-194 \dot{G} \dot{H}+97 \dot{G}^2-132 H^4-72 H^2 \dot{H}+97 \dot{H}^2 \big). \notag
\end{alignat}
\vspace{-0.8cm}
\begin{alignat}{3}
  B_2 &= \tfrac{7168}{675} \big( 2 G^4+38 G^3 H-49 G^2 H^2-46 G^2 \dot{H}+46 \dot{G} G^2-24 G H^3-108 \dot{G} H^2\notag
  \\
  &\quad\, 
  + 62 \dot{G} G H   -62 G H \dot{H}-154 \dot{G} \dot{H}+77 \dot{G}^2+33 H^4+108 H^2 \dot{H}+77 \dot{H}^2 \big).
\end{alignat}
We remark that the equation of motion (\ref{eq:TensorP}) can be derived from this action.
The effective action for $h_+$ takes slightly different form, but we obtain the same one
after substituting the background metric.


\end{document}